\documentclass[10pt,conference]{IEEEtran}

\usepackage{amsmath}
\usepackage{amssymb}
\usepackage{enumerate}
\let\labelindent\relax
\usepackage[shortlabels]{enumitem}

\usepackage{amsthm}
\usepackage{calc}

\newtheorem{definition}{Definition}
\newtheorem{theorem}{Theorem}

\newcommand{\BT}{\begin{theorem}}
\newcommand{\ET}{\end{theorem}}

\newcommand{\BD}{\begin{definition}}
\newcommand{\ED}{\end{definition}}

\newcommand{\BCR}{\begin{corollary}}
\newcommand{\ECR}{\end{corollary}}

\newcommand{\BEX}{\begin{example}}
\newcommand{\EEX}{\end{example}}

\newcommand{\BL}{\begin{lemma}}
\newcommand{\EL}{\end{lemma}}

\newcommand{\BP}{\begin{proposition}}
\newcommand{\EP}{\end{proposition}}

\newcommand{\BCM}{\begin{claim}}
\newcommand{\ECM}{\end{claim}}

\newcommand{\BPF}{\begin{proof}}
\newcommand{\EPF}{\end{proof}}
\newcommand{\BEN}{\begin{enumerate}}
\newcommand{\EEN}{\end{enumerate}}

\newcommand{\BI}{\begin{itemize}}
\newcommand{\EI}{\end{itemize}}

\newcommand{\BO}{\begin{observation}}
\newcommand{\EO}{\end{observation}}

\newcommand{\BDS}{\begin{description}}
\newcommand{\EDS}{\end{description}}

\def\E{\mathop{\mathbb{E}}\displaylimits}   %

\newcommand{\ignore}[1]{}

\newcounter{defcounter}
\newlength{\protowidth}

\newcommand{\olrk}[1]{%
   \ifx\nursymbol#1\else\!\!\mskip4.5mu plus 0.5mu\left(#1\right)\fi}
\newcommand{\elrk}[1]{%
   \ifx\nursymbol#1\else%
        \!\!\mskip4.5mu plus0.5mu\left[\mskip2.5mu plus0.5mu #1\right]\fi}

\renewcommand{\paragraph}[1]{\noindent \textbf{#1}~}
\usepackage{etoolbox}

\def\path{\ensuremath{{\sf location}}}

\usepackage{graphicx}
\graphicspath{{./figures/}}

\usepackage[
  hidelinks,
  pdftitle={ObliviSync},
]{hyperref}

\usepackage[capitalise,noabbrev]{cleveref}

\usepackage{xcolor}

\usepackage{tikz}
\usetikzlibrary{calc,positioning,shapes,decorations.pathreplacing,shadows}

\usepackage{xspace}

\newcommand{\sys}[0]{\ensuremath{{\sf Oblivi\-Sync}}\xspace}
\newcommand{\sysrw}[0]{\ensuremath{{\sf Oblivi\-Sync\mbox{-}RW}}\xspace}
\newcommand{\sysro}[0]{\ensuremath{{\sf Oblivi\-Sync\mbox{-}RO}}\xspace}

\begin{document}

\title{ObliviSync: Practical Oblivious\\ File Backup and Synchronization}

\author{
\IEEEauthorblockN{
Adam J.~Aviv,\quad
Seung Geol Choi,\quad
Travis Mayberry,\quad
Daniel S.~Roche
}
\IEEEauthorblockA{
United States Naval Academy\\
{\tt \{aviv,choi,mayberry,roche\}@usna.edu}} 
}

\maketitle

\begin{abstract}

Oblivious RAM (ORAM) protocols are powerful techniques that hide a client's
data as well as access patterns from untrusted service providers.
  We present an oblivious cloud storage system, ObliviSync, that specifically
targets one of the most widely-used personal cloud storage paradigms:
synchronization and backup services, popular examples of which are Dropbox,
iCloud Drive, and Google Drive.  This setting provides a unique opportunity
because the above privacy properties can be achieved with a simpler form of
ORAM called write-only ORAM, which allows for dramatically increased efficiency
compared to related work.  Our solution is asymptotically optimal and
practically efficient, with a small constant overhead of approximately 4x
compared with non-private file storage, depending only on the total data size
and parameters chosen according to the usage rate, and not on the number or
size of individual files. Our construction also offers protection against
timing-channel attacks, which has not been previously considered in ORAM
protocols. We built and evaluated a full implementation of ObliviSync that
supports multiple simultaneous read-only clients and a single concurrent
read/write client whose edits automatically and seamlessly propagate to the
readers. We show that our system functions under high work loads, with
realistic file size distributions, and with small additional latency (as
compared to a baseline encrypted file system) when paired with Dropbox as the
synchronization service.
  
\end{abstract}

\section{Introduction}
\paragraph{ORAM: security and efficiency.}
ORAM is a protocol which allows a client to access files (commonly
abstracted as $N$ fixed-length \emph{blocks} of data) stored on an
untrusted server in such a way that the server learns neither the
\emph{contents} of files nor the \emph{access patterns} of which files
were accessed at which time(s). This is traditionally accomplished by
doing some type of shuffling on the data in addition to
reading/writing the chosen block.  This shuffling ensures that the
server cannot correlate logical blocks based on their storage
locations.

ORAM is a powerful tool that solves a critical problem in cloud
security.  Consider a hospital which uses cloud storage to backup
their patient records.  Even if the records are properly encrypted, an
untrusted server that observes which patient files are modified will
learn sensitive medical information about those patients.  They will
certainly learn that the patient has visited the hospital recently,
but also may learn things like whether the patient had imaging tests
done based on how large the file is that is updated.  Moreover, they
might learn for instance that a patient has cancer after seeing an
oncologist update their records.  This type of inference, and more,
can be done despite the fact that the records themselves are encrypted
because the \emph{access pattern} to the storage is not hidden.

Unfortunately, in order to achieve this obliviousness ORAMs often
require a substantial amount of shuffling during every access, so much
so that even relatively recent ORAM constructions could induce a
several-thousand-fold overhead on communication
\cite{AC:SCSL11, kushilevitz2012security}.  Even Path ORAM
\cite{CCS:SDSFRY13}, one of the most efficient ORAM constructions
to date, has a practical overhead of 60-80x on moderately sized
databases compared to non-private storage.

\paragraph{The setting: personal cloud storage.}
Our setting consists of an untrusted cloud provider and one or more
clients which backup data to the cloud provider.  If there are
multiple clients, the cloud provider propagates changes made by one
client to all other clients, so that they each have the same version
of the filesystem.  We emphasize that although we may use ``Dropbox''
as a shorthand for the scenario we are addressing, our solution is not
specific to Dropbox and will work with any similar system.  This
setting is particularly interesting for a number of reasons:

\begin{enumerate}
\item It is one of the most popular consumer cloud services used
  today, and is often colloquially synonymous with the term ``cloud.''
  Dropbox alone has over 500 million users \cite{dropboxusers}.

\item The interface for Dropbox and similar storage providers is
  ``agnostic,'' in that it will allow you to store any data as long as
  you put it in the designated synchronization directory.  This allows
  for one solution that works seamlessly with all providers.

\item Synchronization and backup services do not require that the ORAM
  hide a user's read accesses, only the writes. This is because
  (by default) every client stores a \emph{complete local copy} of their
  data, which is synchronized and backed up via communication of changes
  to/from the cloud provider.
\end{enumerate}

\paragraph{Our goal.}
In this paper, we present an efficient solution for oblivious storage
on a personal cloud synchronization/backup provider such as (but not
limited to) Dropbox or Google Drive.

\paragraph{Write-only ORAM.}
The third aspect of our setting above (i.e., we don't need to hide read
accesses) is crucial to the efficiency of our system.  Each client
already has a copy of the database, so when they read from it they do
not need to interact with the cloud provider at all.  If a client
writes to the database, the changes are automatically propagated to
the other clients with no requests necessary on their part.
Therefore, the ORAM protocol only needs to hide the write accesses
done by the clients and not the reads.  This is important because
\cite{CCS:BMNO14} have shown that \emph{write-only} ORAM can be
achieved with optimal asymptotic communication overhead of $O(1)$.  In
practice, write-only ORAM requires only a small constant overhead of
3-6x compared to much higher overheads for fully-functional ORAM
schemes, which asymptotically are $\Omega(\log N)$.

Note that Dropbox (and other cloud providers) do have client software
that allows retrieval of only individual files or directories, for
instance the Google Drive web interface. However, to achieve privacy in
those settings with partial synchronization would require the
full functionality of Oblivious RAM that hides both reads and writes.
We instead specifically target the full-synchronization setting for
two reasons:
\begin{enumerate}
\item It is the default behavior for the desktop clients of Dropbox,
  Google Drive, OneDrive, and others, making it a widely used,
  practical scenario.
\item Read-write ORAMs are subject to a well-known lower bound of
  $\Omega(N \log{N})$\cite{GO96}.  We aim to show that in a
  synchronization setting substantially better performance can be
  achieved that rivals the performance of insecure storage.
\end{enumerate}

\paragraph{Providing cloud-layer transparency.}
One of the most noteworthy aspects of Dropbox-like services is their
ease of use.  Any user can download and install the requisite
software, at which point they have a folder on their system that
``magically'' synchronizes between all their machines, with no
additional setup or interaction from the user.  In order to preserve
this feature as much as possible, our system implements a FUSE
filesystem that mounts on top of the shared directory, providing a new
{\em virtual frontend directory} where the user can put their files to
have them \emph{privately} stored on the cloud.  The FUSE module uses
the original shared directory as a backend by storing blocks as
individual files.  We stress that substantial work is needed to add
{\em support for filesystem features} such as filenames and sizes,
since the storage of ORAMs is traditionally modeled as only a flat
array of $N$ fixed-length blocks indexed by the numbers $[0, N)$.

\paragraph{Supporting variable-size files.}
When addressing the personal cloud setting, a crucial aspect that must
be dealt with is the variable sizes of the files stored in such a
system.  Traditionally, ORAMs are modeled as storage devices on $N$
{\em fixed-length} blocks of data, with the security guarantee being
that any two access patterns of the \emph{same length} are
indistinguishable from each other.  In reality, files stored on
Dropbox are of varying (sometimes unique) lengths.  This means that a
boilerplate ORAM protocol will actually not provide obliviousness in
such a setting because the file size, in multiples of the block size,
will be leaked to the server for every access.  When file sizes are
relatively unique, knowing the size will enable the server to deduce
which individual file is being accessed, or at least substantially
reduce the number of possibilities.  Therefore our solution
additionally includes a mechanism for {\em dynamically batching
  together variable-length files} to hide their size from the
server. Furthermore, our solution is \emph{efficient} as we prove its
cost scales linearly with the total size (and not number) of files
being written, regardless of the file size distribution. 
The batching aspect of our construction also allows us to protect
against \emph{timing-channel attacks} (where the precise time of an access leaks information about it), which are not usually
considered in ORAM protocols.

\paragraph{Summary of our contribution.}
To summarize, our contributions in this paper include:

\begin{enumerate}
\item A complete ORAM system designed for maximum efficiency and
  usability when deployed on a synchronization/backup service like
  Dropbox.

\item A FUSE implementation of these contributions, incorporating
  variable size files as well as important filesystem functionality
  into ORAM including the ability to store file names, resize files
  and more.

\item A proof of strong security from an untrusted cloud provider,
  even considering the timing side-channel.

\item 

Theoretical evaluation showing that the throughput of our scheme requires only
4x bandwidth overhead compared to that of unencrypted and non-private storage,
regardless of the underlying file size distribution%
\footnote{
The bandwidth is actually set directly according to the system
parameters. If it is too high, ``dummy'' write operations are performed
to hide the access pattern. Our system works as long as the bandwidth is
set to 4x higher than the actual amount of data written. Of course,
the user may set the parameters poorly due to not knowing their usage in
advance, in which case the bandwidth may be higher due to the required
dummy writes. See \Cref{sec:sob}.
}
.
We also show that our scheme has very high storage utilization, requiring only
1.5-2.0x storage cost overhead in practice.

\item An empirical evaluation of the system that shows that \sys
  performs better than the theoretic results for both throughput and
  latency, and \sys functions with limited overheads and delays when
  working with Dropbox as the cloud synchronization service.
  
\end{enumerate}

\section{Efficient Obliviousness for Dropbox}

\subsection{Overview of Write-only ORAM}
We start by describing the write-only ORAM of \cite{CCS:BMNO14}, as it
informs our construction.  

\paragraph{The setting.}
To store $N$ blocks in a write-only ORAM, the server holds an array of $2N$
encrypted blocks.  Initially, the $N$ blocks of data are shuffled and stored in random
locations in the $2N$-length array, such that half of the blocks in the array
are ``empty''.  However, every block is encrypted with an IND-CPA encryption
scheme so the server cannot learn which blocks are empty and which are not.  The
client stores a {\em local dictionary} (or sometimes called a position map)
which maps a logical address in the range $(0,N]$ to the location in the server
array where it is currently stored, in the range $(0,2N]$.  Using this
dictionary, the client can find and read any block in the storage that it needs,
but the server will not know the location of any individual block.

\paragraph{Read and write operations.}
Since by definition a write-only ORAM does not need to hide reads, they are
performed trivially by reading the local dictionary and then the corresponding
block from the ORAM.  Write operations, however, require additional work.  When
the client wishes to write a block to the ORAM, it chooses $k$ random locations
in the array out of $2N$, where $k$ is a constant parameter.  With high
probability, at least one out of these $k$ locations will be empty, and
the new block is written into that location while re-encrypting the
other $k-1$ locations to hide which block was changed.
After writing the block, the client also updates their
dictionary to indicate that the block now resides in its new location.  The old
location for this block is implicitly marked empty because no entry in the
dictionary now points to it.

\paragraph{Achieving obliviousness.}
Since every write operation sees the client accessing $k$ randomly chosen blocks
in the ORAM, independent of the logical address of the block that is being
written, it cannot reveal any information about the client's access pattern.
The only situation that can cause the client to reveal something is if the $k$
chosen locations do not contain any free blocks, and it has nowhere to write the
new one.  Since every block has $1/2$ probability of being empty, the chance
that there are no free blocks will be $2^{-k}$, so $k$ can be set to the
security parameter $\lambda$ to give a negligible chance of failure.

\paragraph{Efficiency with stash on the client.}
However, setting $k=\lambda$ actually does not result in $O(1)$ overhead; since
$\lambda > \log{N}$, the overhead is $\Omega(\log N)$. 
On average, the client finds $k/2$ empty blocks during a single write, many more than
are needed.  
If the client instead stores a buffer of blocks that it wants to
write, and writes as many blocks from the buffer as he finds empty blocks, $k$ can be set much more
aggressively.  It is shown in \cite{CCS:BMNO14} that $k=3$ is sufficient to
guarantee with high probability that the stash will never exceed $O(\log{N})$.
This makes the final overhead for write-only ORAM 3x that of non-private
storage. %

\paragraph{Maintaining the dictionary file.}
The final important detail is that the dictionary file requires $O(N \log{N})$
bits of storage, which might be too large for the client to store locally.
Fortunately it is relatively simple to store this dictionary recursively in
another ORAM~\cite{CCS:BMNO14,CCS:SDSFRY13}.  For some block and databases
sizes, however, it might be quite reasonable for the client to store the entire
dictionary itself. 
Jumping ahead, in our system, the client locally stores the dictionary file
(called the {\em filetable}) as an important metadata structure of the entire
file system, in order to keep track of the actual position of each file block.
See the detailed discussion in Section~\ref{sec:system-desc}. 

\subsection{Overview of Our System}

\paragraph{The setting.}
Our \sys system uses the idea of write-only ORAM on top of any file backup or
synchronization tool in order to give multiple clients simultaneous updated
access to the same virtual filesystem, without revealing anything at all to the
cloud service that is performing the synchronization itself, even if the
cloud service is corrupted to become an honest-but-curious adverary. 
Write-only ORAM is ideal for this setting because \emph{each client stores an
entire copy of the data}, so that only the changes (write operations) are
revealed to the synchronization service and thus only the write operations need
to be performed obliviously. 

\paragraph{Improvements over write-only ORAM.}
Compared to the previous write-only ORAM construction~\cite{CCS:BMNO14}, we
make significant advances and improvements to fit this emergent application
space:
\begin{itemize}
  \item \textbf{Usability}: Users interact with the system as though it is
a normal system folder. All the encryption and synchronization happens
automatically and unobtrusively.

\item \textbf{Flexibility}: We support a real filesystem and use innovative
methods to handle variable-sized files and changing client roles (read/write
vs. read-only) to support multiple users.

\item \textbf{Strong obliviousness}: The design of our system not only
provides obliviousness in the traditional sense, but also \emph{protects against
timing channel attacks}. It also conceals the total number of write operations,
a stronger guarantee than previous ORAM protocols. 

\item \textbf{Performance}: Our system well matches the needs of real file
systems and matches the services provided by current cloud synchronization
providers. It can also be tuned to different settings based on the
desired communication rate and delay in synchronization.
\end{itemize}

\begin{figure}[t]
  \centering
  \includegraphics[width=\minof{10 cm}{0.9\linewidth}]{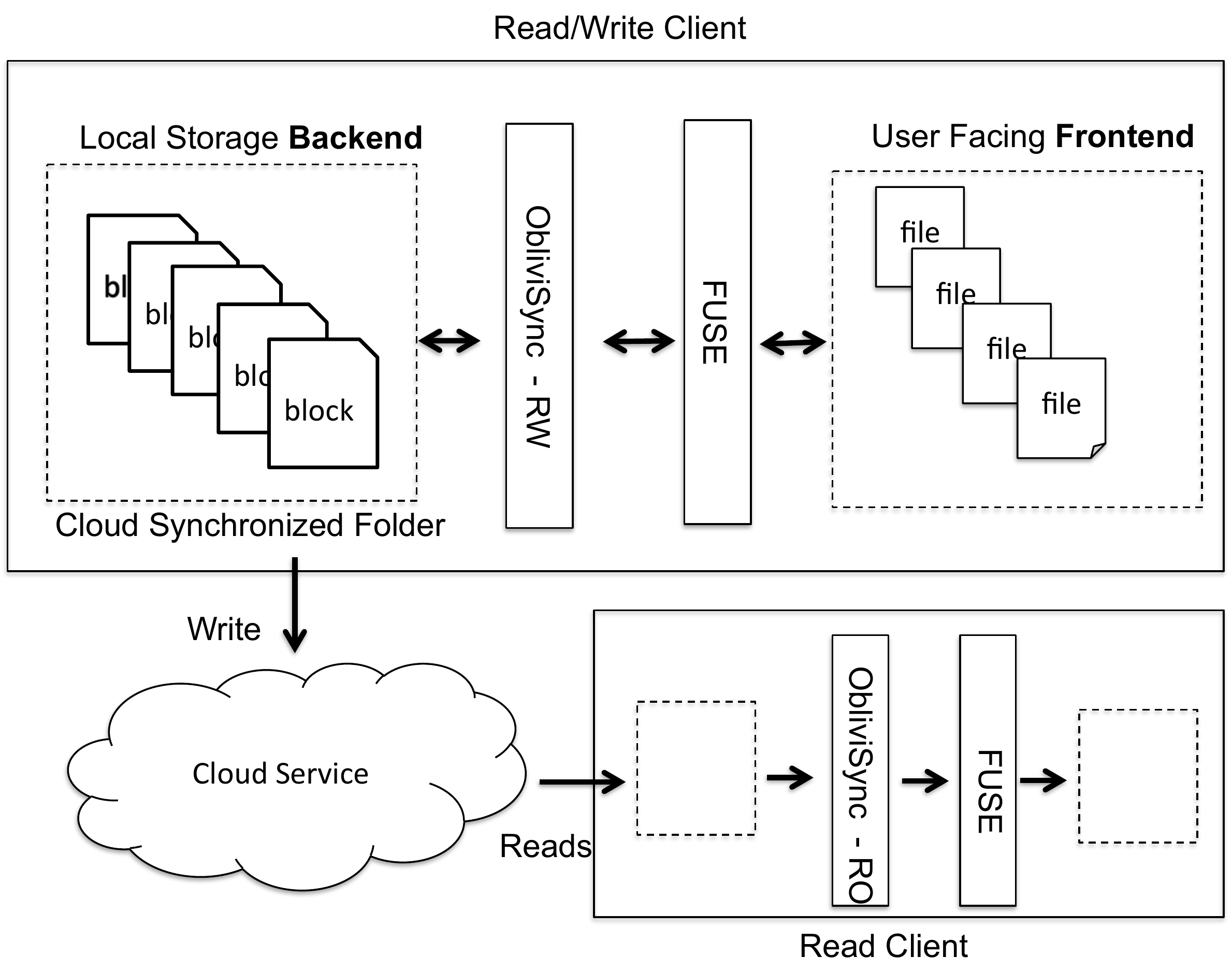}
  \caption{Diagram for \sys}
  \label{fig:sys}
\end{figure}

\paragraph{Basic architecture.}
The high-level design of \sys is presented in \cref{fig:sys}.  There are
two types of clients in our system: a read/write client (\sysrw) and a
read-only client (\sysro). 
At any given time, there can be any number of \sysro's active
as well as zero or one \sysrw clients.
We note that a given device may work as a read-only client in one period of
time and as a write-only client in other periods of time.\footnote{How to make
sure that only one write-only client operates at a given time is out the scope,
and in this paper, we will simply assume the existence of the procedure to
enforce it.} 
Both clients consist of an actual \emph{backend} folder as well as a virtual
\emph{frontend} folder, with a FUSE client running in the background to
seamlessly translate the encrypted data in the backend to the user's view in
the frontend virtual filesystem.

We rely on existing cloud synchronization tools to keep all clients'
backend directories fully synchronized. This directory consists of
encrypted files that are treated as generic storage blocks, and
embedded within these storage blocks is a file system structure
loosely based on i-node style file systems which allows for
variable-sized files to be split and packed into fixed-size
units. Using a shared private key (which could be derived from a password) the job of both clients \sysro and
\sysrw is to decrypt and efficiently fetch data from these encrypted
files in order to serve ordinary read operations from the client
operating in the frontend directory.

The \sysrw client, which will be the only client able to change the
backend files, has additional responsibilities: (1) to maintain the
file system encoding embedded within the blocks, and (2) to perform
updates to the blocks in an oblivious manner using our efficient
modification of the write-only ORAM described in the previous
subsection.

\paragraph{User transparency with FUSE mount.}
From the user's perspective, however, the interaction with the
frontend directory occurs as if interacting with any files on the host
system. This is possible because we also implemented a FUSE mount
(file system in user space) interface which displays the embedded file
system within the backend blocks to the user as if it were any other file
system mount. Under the covers, though, the \sysro or \sysrw clients
are using the backend directory files in order to serve all data
requests by the client, and the \sysrw client is additionally
monitoring for file changes/creations in the FUSE mount and
propagating those changes to the backend.

\paragraph{Strong obliviousness through buffered writes.} \label{sec:sob}
In order to maintain obliviousness, these updates are \emph{not}
immediately written to the backend filesystem by the \sysrw client.
Instead, the process maintains a \emph{buffer} of writes that are staged
to be committed. 
At regular timed intervals, random blocks from the backend are loaded, repacked
with as much data from the buffer as possible, and then re-encrypted and
written back to the backend folder. From there, the user's chosen file
synchronization or backup service will do its work to propagate the changes to
any read-only clients. 
Moreover, even when there are no updates in the buffer, the client pushes dummy
updates by rewriting the chosen blocks with random data. 
In this way, as the number of blocks written at each step is fixed, and these
writes (either real or dummy) occur at regular timed intervals, an adversary
operating at the network layer is unable to determine anything about the file
contents or access patterns.
Without dummy updates, for example, the adversary can make a reasonable guess
about the size of the files that the client writes; continuted updates without
pause is likely to indicate that the client is writing a large file. Note that
in some cases, revealing whether a client stores large files (e.g., movies) may
be sensitive. 
Further details on all of these components can be found in
\cref{sec:details}. The full source code of our implementation is
available on GitHub~\cite{ourgithub}.

\section{Security Definitions}

\subsection{Write-only Oblivious Synchronization}

\paragraph{Block-based filesystem.} 
Our system has more capabilities than a standard ORAM, including support for
additional filesystem operations, so we require a modified security definition
which we present here. We first formally define the syntax of a block-based
filesystem with block size $B$.

\BI
\item $\mathsf{create}$(filename): create a new (empty) file. 
\item $\mathsf{delete}$(filename): remove a file from the system.
\item $\mathsf{resize}$(filename,size): change the size of a file. The
size is given in bytes, and can be greater or smaller than the current
size.
\item $\mathsf{write}$(filename,offset,length): write data to the
    identified file according to the offset and length arguments.
    The offset is a \emph{block offset}. Unless the offset refers to the
    last block in the file, length must be a multiple of $B$.
\item $\mathsf{read}$(filename,offset,length) $\rightarrow\mathsf{data}$:
    read data from the identified file according to the offset and length arguments. 
    Again, offset is a block offset, and length must be a multiple
    of the block size $B$ unless the read includes the last offset.
\EI

For simplicity, we only consider these five core operations. Other
standard filesystem operations can be implemented using these core functionalities.

\paragraph{Obliviousness and more.} The original write-only ORAM definition
in~\cite{CCS:BMNO14} requires indistinguishability between any two write
accesses with {\em same data sizes}. However, the definition does not consider
the time at which write operations take place. Here, we put forward a stronger
security notion for the file system that additionally hides both the data
length and the time of non-read operations.

For example, we want to make sure all the following operation
sequences are indistinguishable: 
\BI[noitemsep]
\item no write operations at all
\item $\mathsf{write}$(file1,1,5) and $\mathsf{write}$(file2,3,3) at time 3, and\\
$\mathsf{write}$(file1,6,9) at time 5
\item $\mathsf{write}$(file2,1,20) at time 5
\EI

For this purpose, we first define $(L, t)$-fsequences. Here, the parameter $L$
is the maximum number of bytes that may be modified, and $t$ is the latest time that is
allowed. For example, the above sequences are all $(20,5)$-fsequences, since
all sequences write at most 20 bytes data in total
and have the last write before or at time $5$.

\BD[\bf $(L,t)$-fsequence] 
A sequence of non-read operations for a block filesystem is a {\sf
$(L,t)$-fsequence} if the total number of bytes to be modified in the
filesystem metadata and file data is at most $L$, and the last operation takes
place before or at time $t$.  
\ED

Our goal is to achieve an efficient block filesystem construction such that any
two $(L,t)$-fsequences are indistinguishable.  

\BD[\bf Write-only strong obliviousness] Let $L$ and $t$ be the
parameters for
fsequences. A block filesystem is {\sf write-only strongly-oblivious with
running time $T$}, if for any two $(L,t)$-fsequences $P_0$ and $P_1$, it holds
that:
\BI
\item The filesystem finishes all the tasks in each fsequence within time T with
probability $1 - neg(\lambda)$, where is $\lambda$ is the security parameter.
\item The access pattern of $P_0$ is computationally indistinguishable to that
of $P_1$. 
\EI
\ED

\section{System Details}
\label{sec:details}

As described previously, the basic design of \sys is presented in
Figure~\ref{fig:sys}. In this section we highlight the implementation details
further. In particular, we describe the implementation components focusing on
interesting design challenges and user settings that can be used to tune the
performance.

\subsection{Filesystem Description}
\label{sec:system-desc}
First, we describe the data organization of the backend files that
form the storage mechanisms for the encrypted filesystem. We note that
our filesystem is \emph{specifically tailored for the \sys use case},
and this design is what leads to our practical performance gains.

\paragraph{Files, fragments, blocks, block-ids.}
The user of an \sys client is creating, writing, reading, and deleting
\emph{logical files} in the frontend filesystem via the FUSE mount. The \sys client, to
store user files, will break down the files into one or more \emph{fragments},
and these fragments are then stored within various encrypted \emph{blocks} in
the backend. 

Blocks are stored in the backend directory in \emph{block pairs} of
exactly two files each. (Note each block pair resides within a single
file in the backend directory, but we avoid the use of the word ``file''
when possible to disambiguate from the logical frontend files, and instead refer
to these as ``block pairs''.)
We explain why we
chose to combine exactly two blocks in each pair later when discussing performance
optimization. 
Note that it is only these encrypted block pairs in the backend
directory which are seen
(and transferred) by the cloud synchronization service.  

While each block has the same size, files stored in the frontend can have
arbitrary sizes. A file fragment can be smaller than a block size, but not
larger. In particular, each file will consist of an ordered list of
fragments, only the last of which may have size smaller than that of a
full block.

While files are indicated by pathnames as in any normal filesystem,
backend blocks are indexed by numeric {\em block-ids}, with numbering
such that the two blocks that make up a block pair are readily
identified. 
See Figure~\ref{fig:filesys} for a simple example. 

\tikzset{
short/.style={draw,rectangle,text height=8pt,text width=9pt,align=center,fill=gray!30},
sshort/.style={draw,rectangle,text height=8pt,text width=5pt,align=center,fill=gray!30},
medium/.style={draw,rectangle,text height=8pt, text width=0.9cm,align=center,fill=gray!30},
twopart/.style={draw,rectangle,text
width=0.9cm,align=center,text centered, rectangle split, rectangle split
parts=2,text=white},
twopartb/.style={draw,rectangle,text width=0.9cm,align=center,text centered, rectangle split, rectangle split
parts=2},
}
\newcommand\Textbox[2]{%
    \parbox[c][\dimexpr#1-7.7pt][c]{0.9cm}{\centering#2}}

\begin{figure}[t]
\centering
\small
\noindent\begin{tikzpicture}[node distance=0]

\def\inode#1#2{%
\node[medium,label=center:{#2}] (#1) {}}
\def\shnode#1#2#3{%
\node[short,right=of #1, label=center:{#3}] (#2) {}}
\def\sshnode#1#2#3{%
\node[sshort,right=of #1, label=center:{#3}] (#2) {}}
\def\lnode#1#2#3{%
\node[medium,right=of #1, label=center:{#3}] (#2) {}}

\def\unode#1#2#3{%
\node[medium,below=10pt of #1, label=center:{#3}] (#2) {}}

\def\inodes#1#2{\node[twopart,below=1.5cm of #1] (#2) {\Textbox{15pt}{~}\nodepart{second}\Textbox{15pt}{~}}}
\def\lnodes#1#2{\node[twopart,right=5pt of #1] (#2) {\Textbox{15pt}{~}\nodepart{second}\Textbox{15pt}{~}}}
\def\lnodeb#1#2#3#4{\node[twopartb,fill=gray!30,right=5pt of #1] (#2) {\Textbox{15pt}{#3}\nodepart{second}\Textbox{15pt}{#4}}}

\def\lnodex#1#2#3{%
\node[medium,text height=8pt, above right =-14pt and 5 pt of #1, label=center:{#3}] (#2) {}}

\def\lnodey#1#2#3{%
\node[medium,text height=8pt, below right =-14pt and 5 pt of #1, label=center:{#3}] (#2) {}}

\def\lnodez#1#2#3{%
\node[short, text height=8pt, above right =-14pt and 5 pt of #1, label=center:{#3}] (#2) {}}

\def\lnodew#1#2#3{%
\node[sshort, text height=8pt, above right =-14pt and 20 pt of #1, label=center:{#3}] (#2) {}}

\inode{a}{$x_1$};
\lnode{a}{b}{$x_2$};
\lnode{b}{c}{$x_3$};
\shnode{c}{d}{$x_4$};
\node[above=0.2cm of c](lx){\bf Frontend};
\node[left=0.2cm of a](lz){file X};

\unode{a}{ya}{$y_1$};
\lnode{ya}{yb}{$y_2$};
\sshnode{yb}{yc}{$y_3$};
\node[left=0.2cm of ya](lz){file Y};

\inodes{lz}{aa};
\lnodeb{aa}{bb}{$x_2$}{$x_3$};
\lnodes{bb}{cc};
\lnodes{cc}{dd};
\lnodes{dd}{ee};
\lnodes{ee}{ff};
\node[below =2.2 cm of lx](ly){\bf Backend};
\node[below =10pt of aa](laa){file 0};
\node[below =10pt of bb](lbb){file 1};
\node[below =10pt of cc](lcc){file 2};
\node[below =10pt of dd](ldd){file 3};
\node[below =10pt of ee](lee){file 4};
\node[below =10pt of ff](lff){file 5};

\lnodex{bb}{aaa}{$x_1$};
\lnodez{dd}{ccc}{$x_4$};
\lnodew{dd}{ccc}{$y_3$};

\lnodeb{cc}{yyy}{$y_1$}{$y_2$};

\draw[thick, ->] (-2.2cm,-7.5em) -- (-1.7cm,-8.5em) node[above left=7pt and -10pt] {block-id=0};
\draw[thick, ->] (-2.2cm,-11.5em) -- (-1.7cm,-10.5em) node[below left=7pt and -10pt] {block-id=1};
\draw[thick, ->] (0cm,-7.5em) -- (0cm,-8.5em) node[above = 7pt] {block-id=2};

\end{tikzpicture}
\caption{An example of two front-end files stored in the backend
    files\label{fig:filesys}.  
    The frontend file X (resp. file Y) consists of 4 (resp. 3) fragments where
    the last fragment is smaller than the block size.  Each fragment is stored
    in a block. Each backend file contains exactly two blocks. Backend blocks
    are indexed by block-ids, starting with 0. For example, the two blocks in
    file 0 has block-ids 0 and 1, respectively, and the block-ids of the blocks
    where fragments $x_1, x_2, x_3, x_4$ are located are $4, 2, 3, 8$
    respectively. Note the small fragments $x_4$ and $y_3$ are located in the
same block with block-id 8.  }
\end{figure}
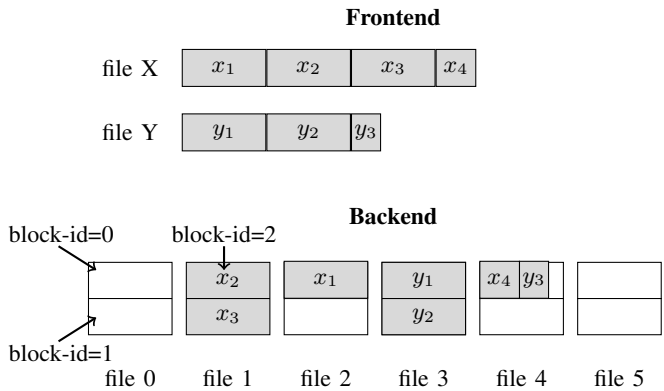

\paragraph{Filetable, file-entries, file-ids.} 
Since a single frontend file consists of a series of fragments (or a single
fragment if the file is small) where each fragment is stored within a block,
\sys needs to keep track of the backend blocks that each file uses so that it
may support file create/update/delete operations effectively.  

For this purpose, \sys maintains a \emph{filetable}, consisting of {\em
file-entries}. Each frontend file is one-to-one mapped to a file-entry, which
maintains some file metadata and a list of block-ids in order to refer to the
blocks that contain the frontend file's fragments, in order. 
In a sense, block-ids (resp., the file-entry) in our system are similar to
position map entries (resp., the position map) in traditional ORAMs. The main
difference is that in order to treat multiple front-end files, our system
maintains a filetable containing multiple file-entries.

The file-entries in the filetable are indexed by {\em file-ids}.  As files
update, the file-id remains constant; however, based on the oblivious writing
procedure, the file fragments may be placed in different backend blocks, so the
block-ids may change accordingly.

\paragraph{Filetable B-tree.}
The filetable mapping file-ids to file-entries is implemented as a
B-tree, with node size B proportional to the size of backend blocks.
In general, the height of this B-tree is $O(\log_B n)$, where $n$ is the
number of files stored. As we will see in \cref{sec:params}, for typical
scenarios the block size is sufficiently large so that the B-tree height
can be at most 1 (i.e., the tree consists of the root node and its
children), and we will assume this case for the remainder.

The leaf nodes are added and stored alongside ordinary files in
blocks. There are two important differences from regular files, however:
leaf nodes are always exactly the size of one full block, and they are
indexed (within the root node) by their block-id directly. This way,
leaf nodes have neither a file-id nor a file-entry.

\paragraph{Directory files.}
As is typical in most filesystems, file paths are grouped into
\emph{directories} (i.e., folders), and each directory contains the
pathnames and file-ids of its contents. Observe that the directory file
only changes when a file within it is created or destroyed, since
file-ids are persistent between file modifications.
Directory files are treated just like any other file, with the special
exception that the \emph{root directory} file is always assigned file-id 0.

\subsection{Design Choices for Performance Optimization}
\paragraph{File-entry cache.}
To avoid frequent writes to the B-tree leaves, we maintain a small
\emph{cache} of recent file-id to file-entry mappings. Like the root
node of the filetable B-tree, the size of this cache is proportional to
the backend block size.

When the cache grows too large, the entries that belong in the
\emph{most common} leaf node among all cache entries are removed and
written to that leaf node. This allows for efficient batching of
leaf node updates and guarantees a significant fraction of the cache is
cleared whenever a leaf node is written.

In fact, if the block size is large enough relative to the number
of files, the cache alone is sufficient to store all file-entries,
and the B-tree effectively has height 0 with no leaf nodes.

\paragraph{Superblock.}
Observe that every time any file is changed, its file-entry must be
updated in the cache, which possibly also causes a leaf node to be
written, which in turn (due to the oblivious shuffling of block-ids on writes)
requires the root node of the filetable to be re-written as well.

Since the root node of the filetable B-tree and the file-entry cache are
both changed on nearly every operation, these are stored specially in a
single designated backend block pair called the \emph{superblock}
that is written on every operation
and never changes location. Because this file is written every time,
it is not necessary to shuffle its location obliviously. 

The superblock also contains a small (fixed) amount of metadata for the
filesystem parameters such as the block size and total number of backend
block pairs. As mentioned
above, the size of both the B-tree root node and the file-entry cache
are set proportionally to the backend block size, so that the superblock
in total has a fixed size corresponding to one block pair; our
implementation stores this always in backend file 0.

\paragraph{Split block: a block with small fragments.}
At the block level, there can be two types of blocks: a {\em full
  block} where the fragment stored within is as large as the block size
and inhabits the entirety of the block, or a {\em split block} where
multiple fragments smaller than the block size are stored within
the same block.  When a large file is stored as a series of fragments,
we maintain that all fragments \emph{except possibly for the last
  fragment} are stored in full blocks. That is, there will be at most
one split-block fragment per file. 

Introducing split blocks allows the system to use the backend storage more
efficiently. For example, without using split blocks, 100 small fragments from
100 frontend files will consume 100 backend blocks, but by placing multiple
small fragments into a single split block, we can reduce the number of blocks
consumed. 

Looking up the data in a full block is straightforward: given the
block-id value, \sys fetches the given block and decrypts its contents.
In addition to the actual data, we also store the file-id of the file
within the block itself as metadata.  This will facilitate an easy
check for whether a given block has become \emph{stale}, as we will
see shortly.

For a split block, however, the system also needs to know the location
of the desired fragment within the block.  The information is stored
within the block itself in the {\em block table} which maps file-ids to
offsets. With the size of the file from the filetable, it is
straightforward to retrieve the relevant data. A full block can then
be simply defined as a block {\em without} a block table, and the
leading bit of each block is used to identify whether the block is
full or split.

\paragraph{Two blocks in a backend file.}
All backend blocks are grouped into \emph{pairs} of two consecutive blocks where
each pair of blocks resides within a single backend file.
Importantly, we relax slightly the indexing requirements so that {\em small fragments are allowed
to reside in either block without changing their block-id}. This tiny block-id
mismatch is easily addressed by looking up both blocks in the corresponding
backend file.

Furthermore, as we will see in the sync operation described later, both blocks
in a given pair are randomly selected to be re-packed and rewritten at the same
time. This additional degree of freedom for small fragments is crucial for
bounding the worst-case performance of the system.

\subsection{Read-Only Client}
A read-only client (\sysro) with access to the shared private key is
able to view the contents of any directory or file in the frontend
filesystem by reading (and decrypting) blocks from the backend, but
cannot create, destroy, or modify any file's contents.

To perform a read operation for any file, given the file-id of that
file (obtained via the directory entry), the \sysro first needs to
obtain the block-id list of the file's fragments to then decrypt the
associated blocks and read the content.  This is accomplished in the
following steps:
\begin{enumerate}
  \item Decrypt and read the superblock. 
  \item Check in the file-entry cache. If found, return the corresponding
  block-id list.
  \item If not found in the cache, search in the filetable via the
    B-tree root (part of the superblock) to find the block-id 
    of the appropriate leaf node in the B-tree.
  \item Decrypt and read the leaf node to find the file-entry for the file in
  question, and return the corresponding block-id list and associated metadata.
\end{enumerate}
Once the block-id list has been loaded, the desired bytes of the file
are loaded by computing the block-id offset according to the block
size, loading and decrypting the block specified by that block-id, and
extracting the data in the block.

Given the file-id, it can be seen from the description above that a
single read operation in \sysro for a single fragment requires loading
and decrypting at most 3 blocks from the backend: 
(1) the superblock, (2) a B-tree leaf node (if not found in file-entry cache), and (3) the block containing
the data. 
In practice, we can cache recently accessed blocks (most notably, the
superblock) for a short period in order to speed up subsequent lookups.

\subsection{Read/Write Client}

The read/write client \sysrw encompasses the same functionality as
\sysro for lookups with the added ability to create, modify, and delete
files.

\paragraph{Pending writes buffer.} 
The additional data structure stored in \sysrw to facilitate these write
operations is the \emph{pending writes buffer} of recent, un-committed changes.
Specifically, this buffer stores a list of $(\text{file-id}, \text{fragment},
\text{timestamp})$ tuples. When the \sysrw encounters a write (or create or
delete) operation, the operation is performed by adding to the buffer.  For
modified files that are larger than a block size, only the fragments of the file
that need updating are placed in the buffer, while for smaller files, the entire
file may reside in the buffer.
During reads, the system first checks the buffer to see if the file-id
is present and otherwise proceeds with the same read process as 
in the \sysro description above.
The main motivation of the buffer is to allow oblivious writing
without compromising usability. The user should not be aware of the
delay between when a write to a file occurs and when the corresponding data is
actually synced to the cloud service provider. The function of the buffer
is similar to that of ``stash'' in normal ORAM constructions.

Interestingly, we note that the buffer also provides considerable
\emph{performance} benefits, by acting as a cache for recently-used
elements. Since the buffer contents are stored in memory un-encrypted,
reading file fragments from the buffer is faster than decrypting and
reading data from the backend storage. The buffer serves a dual
purpose in both enabling obliviousness and increasing practical
efficiency.

\begin{figure*}[t]
\small
\begin{center}
\begin{tabular}{lcccc}
\multicolumn{1}{c}{Action} & Buffer & Backend & \sysrw view & \sysro view \\
\hline
0. (initial) 
  & $\{\}$ & $\{f_1,f_2,f_3\}$ & $[f_1,f_2,f_3]$ & $[f_1,f_2,f_3]$ \\
1. Two blocks updated
  & $\{\mathbf{f_2'},\mathbf{f_3'}\}$ & $\{f_1,f_2,f_3\}$ & $[f_1,\mathbf{f_2'},\mathbf{f_3'}]$ & $[f_1,f_2,f_3]$ \\
2. One block synced
  & $\{\mathbf{f_2'}\}$ & $\{f_1,f_2,f_3,\mathbf{f_3'}\}$ & $[f_1,\mathbf{f_2'},\mathbf{f_3'}]$ & $[f_1,f_2,f_3]$ \\
3. Both blocks synced
  & $\{\}$ & $\{f_1,f_2,f_3,\mathbf{f_3'},\mathbf{f_2'}\}$ & $[f_1,\mathbf{f_2'},\mathbf{f_3'}]$ & $[f_1,\mathbf{f_2'},\mathbf{f_3'}]$ \\
4. Stale data removed
  & $\{\}$ & $\{f_1,\mathbf{f_3'},\mathbf{f_2'}\}$ & $[f_1,\mathbf{f_2'},\mathbf{f_3'}]$ & $[f_1,\mathbf{f_2'},\mathbf{f_3'}]$
\end{tabular}
\caption{Example of ordering and consistency in updating two of three
fragments of a single file. Use of the \emph{shadow filetable} ensures
that the \sysro view is not updated
until all fragments are synced to the backend.\label{fig:shadowex}}
\end{center}
\end{figure*}

\paragraph{Syncing: gradual and periodic clearing of the buffer.}
The buffer within \sysrw must not grow indefinitely. In our system,
the buffer size is kept low through the use of a periodic \emph{sync}
operations wherein the buffer's contents are encrypted and stored in
backend blocks.

Each sync operation is similar to a single write procedure in write-only
ORAM, but instead of being triggered on each write operation,
the sync operation happens on a \emph{fixed timer} basis. We call the
time between subsequent sync operations an \emph{epoch} and define this
parameter as the \emph{drip time} of the \sysrw.

Also, similar to the write-only ORAM, there will be a fixed set of backend files
that are chosen randomly on each sync operation epoch. The number of such
backend files that are rewritten and re-encrypted is defined as the \emph{drip
rate}. Throughout, let $k$ denote the drip rate. We discuss these parameters
further and their impact on performance in \cref{sec:params}. 
Overall, each sync operation proceeds as follows:
\begin{enumerate}
\item Choose $k$ backend block pairs randomly to rewrite and decrypt them.
\item Determine which blocks in the chosen backend files are stale, i.e., containing stale fragments. 
\item Re-pack the empty or stale blocks, clearing fragments from the pending
    writes buffer as much as possible. 
\item Re-encrypt the chosen backend block pairs.
\end{enumerate}
The detailed sync operation is described in the next section. 

\paragraph{Consistency and ordering.}
In order to avoid inconsistency, busy wait, or race conditions, the order of
operations for the sync procedure is very important. For each file fragment that
is successfully cleared from the buffer into the randomly-chosen blocks, there
are three changes that must occur:
\begin{enumerate}
\item The data for the block is physically written to the backend.
\item The fragment is removed from the buffer.
\item The filetable is updated with the new block-id for that fragment.
\end{enumerate}

It is very important that these three changes occur \emph{in that order}, so
that there is no temporary inconsistency in the filesystem.  Moreover, the
\sysrw must \emph{wait until all fragments of a file have been synced} before
updating the file-entry for that file; otherwise there could be inconsistencies
in any \sysro clients.

The consistency is enforced in part by the use of a \emph{shadow
filetable}, which maintains the list of old block-ids for any files that
are currently in the buffer. As long as some fragment of a file is in the
buffer, the entry in the filetable that gets stored with the superblock
(and therefore, the version visible to any read-only client mounts), is
the \emph{most recent completely-synced version}).

An example is depicted in \cref{fig:shadowex}.
Say file $f$ consists of three fragments $[f_1, f_2,
f_3]$, all currently fully synced to the backend. Now say an \sysrw
client updates the last two fragments to $f_2'$ and $f_3'$. It may be
the case that, for example, $f_3'$ is removed from the buffer into some
backend block before $f_2'$ is. (This could easily happen because
$f_3'$ is a small fragment that can be stored within a split block,
whereas $f_2'$ is a full block.) At this point, the \emph{shadow
filetable} will still store the location of $f_3$ and not $f_3'$, so
that any \sysro clients have a consistent view of the file. It is only
after all fragments of $f$ are removed from the buffer that the
filetable is updated accordingly.

One consequence is that there may be some duplicate versions of the same
fragment stored in the backend simultaneously, with neither being stale
(as in $f_3$ and $f_3'$ after step 2 in \cref{fig:shadowex}). 
This adds a small storage overhead to the system, but the benefit is that both
types of clients have a clean, consistent (though possibly temporarily
outdated) view of the filesystem. Note that, even in the worst case, no
non-stale fragment is ever duplicated more than once in the backend.

\subsection{Detailed Description of Buffer Syncing}

\paragraph{Step 1: Choosing which blocks to rewrite.}
As in write-only ORAM, $k$ random backend files are chosen to be rewritten at
every sync operation with the following differences: 
\begin{itemize}
\item Each backend file contains a pair of blocks, which implies that $k$ random
    \emph{pairs} of blocks are to be rewritten.

\item In addition, the backend file containing the superblock is \emph{always}
    rewritten.
\end{itemize}
Choosing pairs of blocks together is crucial, since as we have mentioned above,
small fragments are free to move between either block in a pair without changing
their block-ids.  In addition, the superblock must be rewritten on each sync
because it contains the filetable which may change whenever other content is
rewritten to the backend. 

\paragraph{Step 2: Determining staleness.}
Once the blocks to be rewritten are randomly selected and decrypted,
the next task is to inspect the fragments within the blocks to
determine which are ``stale'' and can be overwritten.

Tracking fragment freshness is vital to the system because of the design of
write-only ORAM. As random blocks are written at each stage, modified fragments
are written to {\em new} blocks, and the file-entry is updated accordingly, 
but the stale data fragment is {\em not} rewritten and will persist in the old
block because that old block may not have been selected in this current sync
procedure. Efficiently identifying which fragments are stale becomes crucial to
clearing the buffer. 

A natural, but flawed, solution to tracking stale fragments is to
maintain a bit in each block to mark which fragments are fresh or
stale.  This solution cannot be achieved for the same reason that
stale data cannot be immediately deleted --- updating blocks that are
not selected in the sync procedure are not possible. 

Instead, recall from the block design that each block also stores the file-id for
each fragment. To identify a stale fragment, the sync procedure looks up each
fragment's file-id to get its block-id list. If the current block's identifier is
not in the block-id list, then that fragment must be stale.

\paragraph{Step 3: Re-packing the blocks.}
Then next step after identifying blocks and fragments within those blocks that
are stale (or empty) is to \emph{re-pack} the block with the non-stale fragments
residing within the block and fragments from the buffer.

One important aspect to consider when re-packing the blocks is to address the
fragmentation problem, that is, to reduce the number of blocks that small
fragments use so that there remain a sufficient number of blocks for full-block
fragments. 

A na\"ive approach would be to evict all the non-stale fragments from the
selected blocks and consider all the fragments in the buffer and the evicted
fragments to re-pack the selected blocks with the least amount of internal
fragmentation. While this would be a reasonable protocol for some file systems
to reduce fragmentation, this would require (potentially) changing all of the
file-entries (in particular, the block-id list) for all fragments within the
selected blocks. That would be problematic because it is precisely these old
entries which are likely not to be in the file-entry cache, and therefore doing
this protocol would require potentially changing many filetable B-tree nodes at
each step, something that should be avoided as writes are expensive. 

Instead, we take a different approach in order to address the fragmentation
problem and minimize the block-id updates for the non-stale fragments at the same
time. Our approach has two stages: the placement of non-stale fragments and then
clearing the fragments in the buffer. 

\paragraph{\it Placement of non-stale fragments.}
We use the following rule when addressing the existing non-stale fragments.
\begin{quotation}
\noindent 
Non-stale full-block fragments stay as they are, but \em non-stale small
fragments may move to the other block in the same backend file.
\end{quotation}

Recall that blocks are paired in order and share a single backend file, and so
this flexibility enables small fragments to be re-packed across two blocks to
reduce fragmentation without having to update the block-id value. 
Further, this solution also avoids a ``full-block starvation'' issue in which
all blocks contained just a small split-block fragment. After re-packing, the
small fragments in each block pair may be combined into a single split block,
leaving the other block in the pair empty and ready to store full block
fragments from the buffer. In other words, the re-pack procedure ensures that
existing full-block fragments do not move, but existing small-block fragments
are packed efficiently within one of the blocks in a pair to leave (potentially)
more fully empty blocks available to be rewritten.

\paragraph{\it Pushing fresh fragments.}
At this point, the randomly-chosen blocks are each either: (a) empty,
(b) filled with an existing full-block fragment, or (c) partially
filled with some small fragments.  
The block re-packing first considers any directory file fragments in the
buffer, followed by any regular file fragments, each in FIFO order.
(Giving priority status to directory files is important to maintain low
latency, as discussed in the next section.)
The synchronization process then proceeds the same for all
fragments in the buffer: for each fragment, it tries to
pack it into the randomly-selected blocks as follows:

\begin{itemize}
\item If it is a full-block fragment, it is placed in the first
available empty block (case (a)), if any remain.
\item If it is a small fragment, it is placed if possible in
the first available split block (case (c)) where there is sufficient
room.
\item If it is a small fragment but it cannot fit in any existing split block,
    then the first available empty block (case (a)), if any, is
    initialized as a new split block containing just that fragment.
\end{itemize}

In this way, every buffer fragment is considered for re-packing in order of age,
but not all may actually be re-packed. Those that are re-packed will be removed
from the buffer and their filetable entries will be updated according to the
chosen block's block-id.

A \emph{key observation} that will be important
in our runtime proof later is that after re-packing, either (1) the buffer is
completely cleared, or (2) all the chosen blocks are nonempty.

\paragraph{Step 4: Re-encrypting the chosen backend files.}
After the re-packing is complete, the sync procedure re-encrypts the
superblock (which always goes at index 0), as well as all the
re-packed blocks, and stages them for writing back to
backend files. The actual writing is done all at once, on the timer,
immediately before the next sync operations, so
as not to reveal how long each sync procedure took to complete.

\subsection{Frontend FUSE Mounts}
The FUSE (file system in user space) mounts are the primary entry
point for all user applications. FUSE enables the capture of system
calls associated with I/O, and for those calls to be handled by an
identified process. The result is that a generic file system mount is
presented to the user, but all accesses to that file system are handled
by either the \sysrw or \sysro client that is running in the background.

The key operations that are captured by the FUSE mount and translated into
\sysrw or \sysro calls are as follows:
\begin{itemize}
  \item $\mathsf{create}(filename)$: create a new (empty) file in the
    system in two steps. First a new file-id is chosen, and the corresponding
    file-entry is added to the filetable. Then that file-id is also stored within
    the parent directory file.
  \item $\mathsf{delete}(filename)$: remove a file from the system by removing
      it from the current directory file and removing the associated file-entry
      from the filetable. 
  \item
    $\mathsf{read}(filename,offset,length)\rightarrow\mathsf{data}$ :
    read data from the identified file by looking up its file-id in
    the current directory and requesting the backend \sysrw or \sysro to
    perform a read operation over the appropriate blocks.
  \item $\mathsf{write}(filename,offset,length)$ : write data to the
    identified file by looking up its file-id in the current directory
    and then adding the corresponding fragment(s) to the \sysrw's
    buffer for eventual syncing.
  \item $\mathsf{resize}(filename,size)$ : change the size of a file
    by looking up its file-entry in the current directory and changing the
    associated metadata. This may also add or remove entries from the
    corresponding block-id list if the given size represents a change in
    the number of blocks for that file. Any added blocks will have negative
    block-id values to indicate that the data is not yet available.
\end{itemize}
Of course, there are more system calls for files than these, such as
{\tt open()} or {\tt stat()} that are implemented within the FUSE
mount, but these functions succinctly encompass all major
operations between the frontend FUSE mount and backend file system
maintenance.

As noted before, the FUSE mount also maintains the file system's
directory structure whose main purpose is to link file names to their
file-id values, as well as store other expected file statistics. The
directory files are themselves treated just like any other file, except
that (1) the root directory always has file-id 0 so it can be found on a
fresh (re-)mount, and (2) directory files are given priority when
syncing to the backend.

For a file to become available to the user, it must both be present in
the backend and have an entry in the directory file. Without
prioritization of directory files, it may be the case that some file is
available without the directory entry update, thus delaying access to
the user. Conversely, the opposite can also be true: the directory
entry shows a file that is not completely synchronized. Fortunately,
this situation is easy to detect upon {\tt open()} and an IO-error can
be returned which should be handled already by application making the
system call.

The FUSE module is aware of some backend settings to improve performance,
notably the block size.  When a file is modified, it is tracked by the FUSE
module, but for large files, with knowledge of the block size, the FUSE module
can identify which full fragments of that file are modified and which remain
unchanged. Only the fragments with actual changes are inserted into the
buffer to be re-packed and synced to the backend.

\subsection{Key parameter settings}\label{sec:params}
The tunable parameters for a \sys implementation consist of:

\begin{itemize}
\item $B$: the size of each backend file (i.e., block pair)
\item $N$: the total number of backend files
\item $n$: the total number of frontend files (i.e., logical files) 
\item $t$: the drip time 
\item $k$: the drip rate
\end{itemize}

\paragraph{Backend files.}
The first two parameters $B$ and $N$ depend on the backend cloud service. 
A typical example of such parameters can be taken from the popular Dropbox
service, which optimally handles data in files of size 4MB, so that $B=2^{22}$
\cite{Dra12}, and the maximal total storage for a paid ``Dropbox Pro'' account
is 1TB, meaning $N=2^{18}$ files would be the limit. Note that, as our
construction always writes blocks in pairs, each block pair is stored in a
single file and the block size in \sys will be $B/2$.

\paragraph{Frontend files.}
The next parameter $n$ is not really a parameter \emph{per se} but rather a
limitation, as our construction requires $n \le B^2$ in order to ensure the
filetable's B-tree has height at most 1. For $B=2^{22}$, this means the user
is ``limited'' to roughly 16 trillion files.

\paragraph{Drip time and drip rate.}
The drip time and drip rate are important parameters for the buffer syncing.
The {\em drip time} is the length of the epoch, i.e., the time between two
consecutive syncs to the backend.  The {\em drip rate} refers to how many block
pairs are randomly selected for rewriting on each epoch.

These two parameters provide a trade-off between latency and
throughput. Given a fixed bandwidth limitation of, say, $x$ bytes per second,
$(k+1)B$ bytes will be written every $t$ seconds for $k$ randomly chosen backend
files and the superblock, so that we must have $(k+1)B/t \le x$. Increasing the
drip time and drip rate will increase latency (the delay between a write in the
\sysrw appearing to \sysro clients), but it will increase throughput as the
constant overhead of syncing the superblock happens less frequently.  

We will consider in our experimentation section (see \cref{sec:exp}) the
throughput, latency, and buffer size of the system under various drip rate and
drip time choices. Our experiment indicates that for most uses, the smallest
possible drip time $t$ that allows a drip rate of $k\ge 3$ files per epoch
should be chosen.

\section{Analysis}

\subsection{Time to write all files}

In this subsection we will prove the main \cref{thm:runtime} that
shows the relationship between the number of sync operations, the drip
rate, and the size of the buffer. Recall from the preceding subsection
the parameters $B$ (block pair size), $N$ (number of backend block pairs), and $k$
(drip rate). Specifically, we will show that, with
high probability, a buffer with size $s$ is completely cleared and
synced to the backend after $O\left(\tfrac{s}{Bk}\right)$
sync operations. This is optimal up to constant factors, since only $Bk$
bytes are actually written during each sync.

\begin{theorem}\label{thm:runtime}
For a running \sysrw client with parameters $B, N, k$ as above,
let $m$ be the total size (in bytes) of all non-stale data currently
stored in the backend, and let $s$ be the total size (in bytes) of
pending write operations in the buffer, and suppose that
$m + s \le NB/4$.

Then the expected number of sync operations until the buffer is entirely
cleared is at most $4s/(Bk)$.

Moreover, the probability that the buffer is \emph{not} entirely cleared
after at least 
$\frac{48s}{Bk} + 18r$ 
sync operations is at most $\exp(-r)$.
\end{theorem}

Before giving the proof, let us summarize what the this theorem means
specifically. 

First, the condition $m+s \le NB/4$ means that the
guarantees hold only when at most 25\% of the total backend
capacity is utilized. For example, if using Dropbox
with 1TB of available storage, the user should store at most
250GB of files in the frontend filesystem in order to guarantee the
performance specified in \cref{thm:runtime}.

Second, as mentioned already, the expected number of sync operations is
optimal (up to constant factors), as the total amount of data written in
the frontend cannot possibly exceed the amount of data being written to
the backend.

In the number of syncs $48s/(Bk)+18r$ required to clear the buffer with
high probability, one can think of the parameter $r$ as the number of
``extra'' sync operations required to be \emph{very sure} that the
buffer is cleared. In practice, $r$ will be set proportionally to the
security parameter. A benefit of our construction compared to many other
ORAM schemes is that the performance degradation in terms of the
security parameter is \emph{additive} and not multiplicative. Put
another way, if it takes 1 extra minute of syncing, after all operations
are complete, in order to ensure
high security, that extra minute is fixed regardless of how long the
\sysrw has been running or how much total data has been written.

Finally, a key observation of this theorem is that it does \emph{not} depend on
the distribution of file sizes stored in the frontend filesystem, or
their access patterns, but only the total size of data being stored.
The performance guarantees of our system therefore allow arbitrary
workloads by the user, provided they can tolerate a constant-factor increase in
the backend storage size.

We now proceed with the proof of \cref{thm:runtime}.

\begin{proof}
There are $N$ blocks of backend storage. Each stores some combination of
at most two split blocks and full blocks. Full blocks have size $\tfrac{B}{2}$
each, and split blocks contain multiple fragments summing to size at
most $\tfrac{B}{2}$ each.

Suppose some sync operation occurs (selecting $k$ block pairs from the
backend, removing stale data and re-packing with new fragments from the
buffer), and afterwards the buffer is still not empty. 
Then it \emph{must} be that case that the $k$ block pairs that were
written are at least half filled, i.e.,
their total size is now at least $\tfrac{kB}{2}$. The reason is, if any
block pair
had size less than $\tfrac{B}{2}$, then it could have fit something more (either
a full block or a fragment) from the buffer. But since the buffer was not
emptied, there were no entirely empty blocks among the $k$ block pairs. 

Furthermore, because $m < \tfrac{NB}{4}$ while the buffer is not empty, the
\emph{expected size} of a single, randomly-chosen pair of blocks is less than
$\tfrac{B}{4}$.
By linearity of expectation, the expected size of k randomly selected
block pairs is less than $\tfrac{kB}{4}$.

Combining the conclusions from the preceding paragraphs
we see that, on any sync operation that does not empty
the buffer completely, the $k$ randomly selected block pairs go from expected size
less than $\tfrac{kB}{4}$, to guaranteed size greater than $\tfrac{kB}{2}$. This means the
expected \emph{decrease in buffer size} in each sync is at least
$\tfrac{kB}{4}$.
Starting with s bytes in the buffer, the expected number of syncs is
therefore less than
$\frac{s}{kB/4} = \frac{4s}{Bk}.$

Now we extend this argument to get a tail bound on the probability that
the buffer is not emptied after $T = 48s/(Bk) + 18r$ sync operations, for some
$r \ge 0$.

Call a sync operation \emph{productive} if it results in either the
buffer being cleared entirely, or the size of the buffer decreasing by
at least $\tfrac{kB}{8}$. And define $Y$ to be a random variable for the
total size of the $k$ randomly-selected blocks for a single operation.

From above, a sync is guaranteed to be productive whenever
$Y \le \tfrac{3kB}{8}$ (because the total size after the sync will be at
least $\tfrac{kB}{2}$). We also know from above that $\E[Y] =
\tfrac{kB}{4}$. Using the Markov inequality, we have

$$\Pr\left[Y \le \frac{3kB}{8}\right]
= 1 -  \Pr\left[Y > \frac{3kB}{8}\right]
\ge 1 - \frac{kB/4}{3kB/8} = \frac{1}{3}.$$

That is, each sync is productive with probability at least
$\tfrac{1}{3}$.

Next, define a random variable $X$ to be the number of productive syncs among a series of
$T = 48s/(Bk) + 18r$ sync operations. 
Importantly, if $X \ge 8s/(Bk)$, then the buffer will be cleared at
the end of $T$ syncs.

We see that $X$ is the sum of $T$
i.i.d. Bernoulli trials, each with probability $p=\tfrac{1}{3}$.
Therefore the Hoeffding bound from \cite{Hoe63} tells us that, for any
$\epsilon > 0$,
$$\Pr\left[X < \left(\tfrac{1}{3}-\epsilon\right)T\right] \le
\exp(-2\epsilon^2 T).$$
Setting $\epsilon = \tfrac{1}{6}$ works to bound the probability that
$X < 8s/(Bk)$, since
$\frac{8s}{Bk} < \frac{1}{6}\left(\frac{48s}{Bk} + 18r\right)$
for any $r>0$.
The theorem follows from 
$\displaystyle\exp(-2\epsilon^2 T) = \exp\left(-\tfrac{1}{18}\cdot
\left(\tfrac{48s}{Bk} + 18r\right)\right) < \exp(-r).$
\end{proof}

\subsection{Security}

\BT
Let $\lambda$ be the security parameter. 
Consider \sysrw with parameters $B, N, k$ as above, and with \emph{drip time}
$t$. For any $L$ and $t$ as
fsequence parameters, \sysrw is strongly-secure write-only filesystem with
running time $T = t + \frac{48Lt}{Bk} + 18\lambda t$. 
\ET

\begin{proof}
We need to show the following:
\BI
\item \sysrw finishes all tasks in each sequence within time T with probability
$1 - neg(\lambda)$.
\item For any two $(L, t)$-fsequences $P_0$ and $P_1$, both access
patterns are computationally indistinguishable from each other.
\EI

The first condition is achieved according to \cref{thm:runtime}, since
one sync operation occurs every $t$ seconds.

It is left to show that the second condition also holds. Obliviousness mostly
follows from the original write-only ORAM security. To achieve strong
obliviousness, we stress that \sysrw ~{\em always} writes encrypted data in $k$
files at the backend chosen independently and uniformly at random.  In
particular: 
\BI
\item If there is too much data to be synchronized, the remaining data is
safely stored in the temporary buffer so that the data will be
eventually synchronized. \cref{thm:runtime} makes sure this must happen with overwhelming probability.
\item If there is too little (or even no) data to be
synchronized, the system generates what amounts to dummy traffic
(re-packing the $k$ chosen block pairs with the same data they stored
before).
\EI
Therefore, the second condition is also satisfied. 
\end{proof}

There is an important assumption in both theorems above, namely that the
client is actually \emph{able} to execute the sync operation with the
given drip rate $k$ within each epoch with drip time $t$. If the
parameters $k$ and $t$ are set too aggressively, it may be the case that
the sync operation takes longer than $t$ seconds to complete, and this
fact would be noticed by the cloud server or indeed any network
observer. While the leakage in such cases is relatively minor (an
indication that, perhaps, the client's system load is higher than
normal), it is nonetheless important for security to ensure $t$ and $k$
are set appropriately to avoid this situation of a sync operation
``blowing through'' the epoch.

\section{Experiments}
\label{sec:exp}

We fully implemented \sys using python3 and {\tt \small fusepy}~\cite{fusepy},
and the source code of our implementation is available on GitHub as well as
a video demonstration~\cite{ourgithub} and a Dockerfile for quick setup and testing. 

To evaluate \sys, we performed a series of experiments to test its performance
at the extremes. In particular, we are interested in the following properties:

\begin{itemize}
\item {\em Throughput with fixed-size files}: If the user of \sys were
  to insert a large number of files {\em all at once}, the buffer will
  immediately be filled to hold all the insertions.  How long does it
  take (in number of epochs) for each of the files to sync to the read
  end? 

  \medskip
\item {\em Throughput with variable-size files}: If the users were to
  insert a large number of variable size files all at once, instead of
  fixed-sized ones, how does the throughput change? 
  This experiment will verify Theorem~\ref{thm:runtime}, which states that the
  performance of our system depends on {\em the total number of bytes (instead
  of the total number of files)} in the pending write buffer.

  \medskip
  
\item {\em Latency}: If the user of \sysrw were to insert a large number
  of files {\em one at a time}, how long does it take for each of the
  files to appear to a different \sysro user?

  \medskip
  
\item {\em The size of pending writes buffer}: We also investigate how
  much space the pending writes buffer uses while the system is
  working under different loads with realistic file sizes. Recall the pending writes
  buffer works similarly as the stash in the write-only ORAM, and it
  is important that this buffer does not grow too large. 

  \medskip
  
\item {\em Functionality with Dropbox backend}: Finally, we performed throughput
    and latency experiments with Dropbox as the backend storage mechanism for
    \sys, and compare its performance to that of EncFS~\cite{encfs} on Dropbox.
    EncFS is a per-file encrypted file system tool that provides no
    obliviousness.

\end{itemize}

\subsection{Throughput with Fixed-size Files}
We first consider the throughput in our system. In particular, we are interested
in the performance as it relates to the availability of backend blocks.

To limit the factors in these experiments, we use the following parameters:
\begin{itemize}
\item $N$ = 1024; we used $1024$ backend files (i.e., block pairs). 
\item $n$ = 920; we attempted to insert $920$ frontend files (or 90\% full). 
\item $B$ = 1 MB; each backend file is 1MB in size. 
\end{itemize}

In our experiments, each frontend file is 1MB and thus fills up two full blocks
(including any metadata). There is also two additional block pairs in the system
for the superblock and the directory entry.  Overall, the entire backend storage
for the system was $N \cdot B$ = 1GB.

In the throughput experiment, we established an empty \sysrw and
attached an \sysro to the backend. We then wrote 920 two-block size
files {\em all at once} to the \sysrw FUSE mounted frontend. We then
manually called the synchronize operation that performs the oblivious
writing to the backend. By monitoring the \sysro FUSE mounted
frontend, we can measure how many epochs are required to fully synchronize
all the files.

\begin{figure}[t]
  \centering
  \includegraphics[width=\minof{10 cm}{0.95\linewidth}]{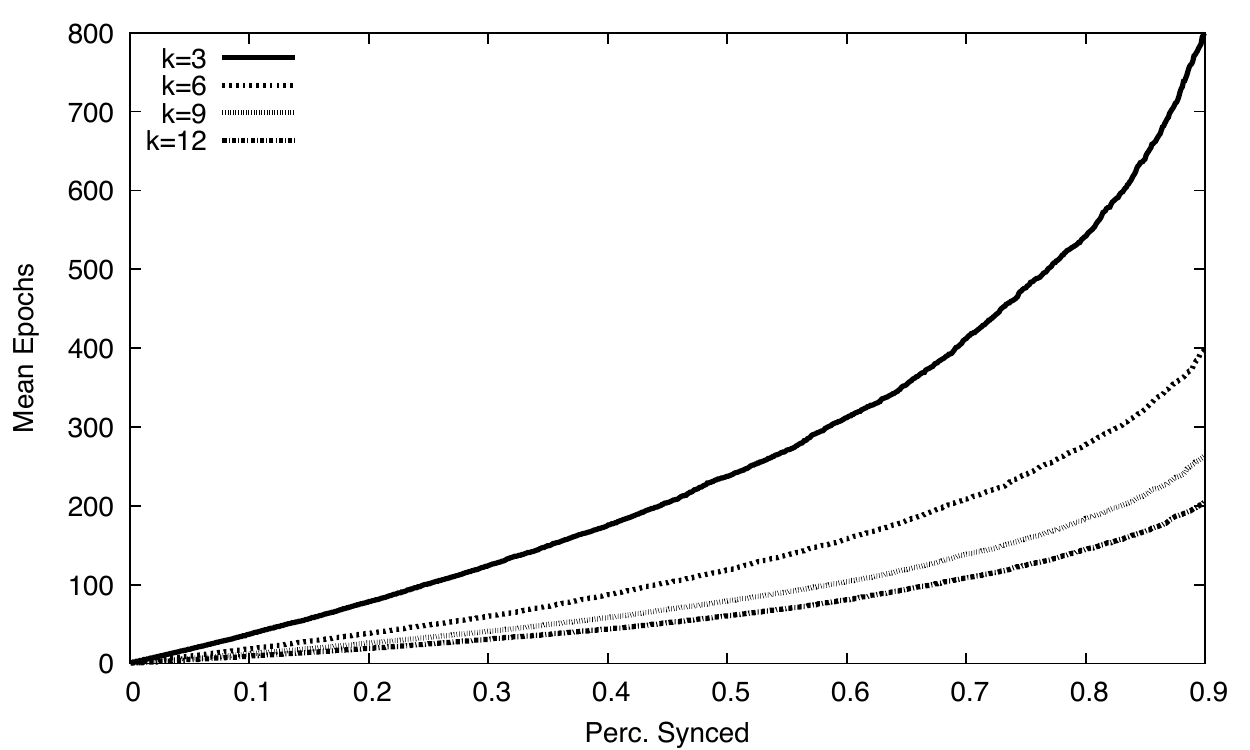}
  \caption{Throughput for different drip rates.  We used $1024$
    backend files, each with 1MB, and attempted to insert $920$
    frontend files all at once, where each frontend file is also
    1MB. The results are the mean of three runs. With drip rate 3 (the
    solid line for $k$ = 3), it takes about $120$ epochs on
    average to sync 25\% of the frontend files. In addition, the
    graph shows the situation shifts as the backend files become more
    full and it becomes harder to clear the buffer.}
  \label{fig:throughput}
\end{figure}

\paragraph{Bandwidth overhead: 2x until 25\% of the load.} 
In Figure~\ref{fig:throughput}, we graph the number of epochs (i.e., the number
of timed sync operations) it takes for that percentage of files to synchronize
and appear at the read-end. We conducted the experiments for different drip
rates ($k$), i.e., the number of backend files randomly selected at each epoch
for writing. The results presented are the average of three runs for drip rates
set to 3, 6, 9, and 12.

As one interesting data point, the graph shows that with drip rate 3 (the solid
line for $k$ = 3), it takes about $\sim 120$ epochs on average to sync 25\% of
the frontend files. Note that the number of bytes that would be transferred to
the cloud storage during 120 epochs is $120 \cdot (k+1) \cdot B$ = 480 MB, and
25\% of the frontend files amounts to $250~\mbox{MB}$. 
So, the experiment shows that {\em the system needs only 2x bandwidth overhead,
when the front-end files occupies at most 25\% of the total cloud storage,}
with the parameters chosen in this experiment. 
This is better performance than what is shown in Thoerem~\ref{thm:runtime}, which
provably guarantees 4x bandwidth overhead. 

\paragraph{Linear costs until 33\% of the load.}
Looking closely at the graph, particularly $k=3$ trend-line, there are
two regimes: linear and super-linear synchronization. In the linear
regime there are enough empty blocks that on each epoch, progress is
very likely to be made by clearing files from the buffer and writing
new blocks to the backend. In the super-linear regime, however, there
are no longer enough empty blocks to reliably clear fragments from the
buffer. For $k=3$, this regime seems to take over around
40\%$\sim$60\%, and the trend-line's slope begins to increase
dramatically. This is because each additional block written further
exacerbates the problem, so it takes an increasing amount of time to
find the next empty block to clear the buffer further.

The inflection point, between linear and super-linear, is particularly
interesting. Apparent immediately is the fact that the inflection
point is well beyond the 25\% theoretic bound; even for a drip rate of
$k=3$, it manages to get at least 1/3 full before super-linear tendencies
take over. Further, notice that for higher drip rates, the inflection
point occurs for higher percentage of fullness for the backend. This
is to be expected; if you are selecting more blocks per epoch, you are
more likely to find an empty block to clear the buffer. But we hasten
to point out that there is a trade-off in practice here.

\begin{figure}[t]
  \centering
  \includegraphics[width=\minof{10 cm}{0.95\linewidth}]{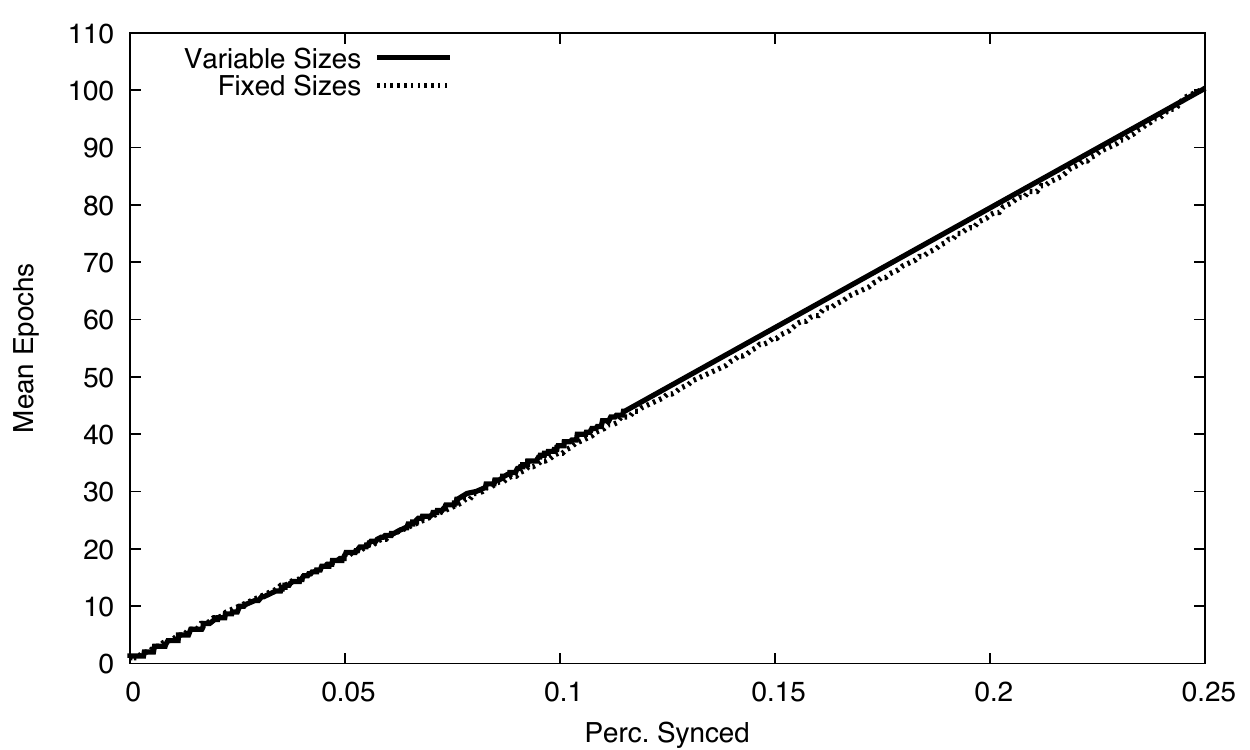}
  \caption{Comparing throughput of inserting realistic workload of variable
      sizes files to the same sized insert of fixed size files. The experiment
      were performed with parameters $N=1024, B = 1\mbox{MB}, k=3$. In the set
      of fixed-size files, each frontend file is 1 MB. There are 4,179 files in
      the set of variable-size files. Both set is 250 MB in total. For both
      sets, the system takes about 100 epochs to sync the 250 MB of frontend
      data. 
  \label{fig:fsizes}}
\end{figure}

\subsection{Throughput with Variable-size Files}
As mentioned above, an important performance property of \sys is that the rate
of synchronization is dependent on the total number of bytes in the pending
write buffer and the fullness of the backend blocks: {\em it does not depend on
the sizes of the individual files.}

To show this property clearly, we performed a similar throughput experiment as
described previously, i.e., with $N = 1024$, $B = 1 \mbox{MB}$, and $k=3$, except we inserted
variable size files that are drawn from realistic file distributions
\cite{Dow01,THB06}: 
\begin{itemize}
\item The variable size files, in total, were 0.25GB, the same size as the fixed
    size files.
\item The files contained 4,179 files, much larger than 250 for the case of
    fixed-size files. 
\item Interestingly, one of the variable size files was significantly larger,
    144MB, roughly the length of a short, TV-episode video.
\end{itemize}

As in the prior throughput experiment, we connected an \sysrw and \sysro to a
shared backend directory.  We loaded the file set completely into the \sysrw,
and then counted how many epochs it takes for the data to appear in the \sysro
FUSE file.

\paragraph{Good performance for variable-size files.}
The primary result is presented in Figure~\ref{fig:fsizes}. The two trend-lines
are nearly identical, and in fact, after three runs, the average number of
epochs needed to synchronize the two file loads is the same, 100 epochs. {\em
This clearly shows that our systems is dependent on the total number of
bytes to synchronize and not the size of the individual files.}

\begin{figure}[t]
  \centering
  \includegraphics[width=\minof{10 cm}{0.95\linewidth}]{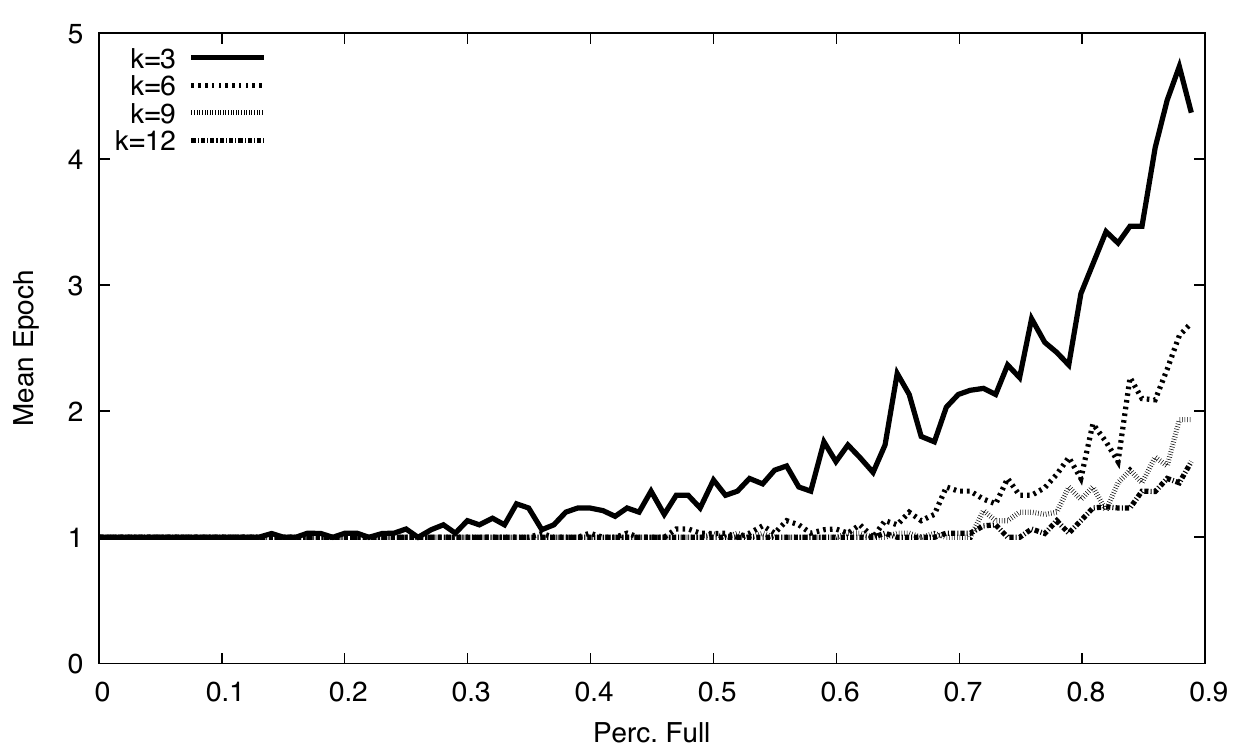}
  \caption{Latency for different drip rates.
   The experiment
      were performed with parameters $N=1024, B = 1\mbox{MB}$. The drip rates
      $k$ varies with $k=3, 6, 9, 12$. The 920 frontend files, each with 1 MB,
      were written one after another. For example, the point (0.7, 2) of the
      solid line with $k=3$ means that once 700 files have been written, it took
      about 2 epochs to sync the 701st file. 
  }
  \label{fig:latency}
\end{figure}

\subsection{Latency}

In this experiment, we are interested in the latency of our system. 
As before, we performed the experiment with $N=1024$ and $B=1\mbox{MB}$, and we
had \sysrw and \sysro clients with a shared backend, writing to \sysrw FUSE
mount and measuring synchronization rate to the \sysro FUSE mount. 
To measure latency, we only add \emph{one frontend file at a time}. For example,
the second frontend file gets written right after the first frontend file is
completely synced. We measured how long it took for each file to synchronize in
terms of the number of manual synchronizations (or epochs) required. 
Again, we varied the drip rate.

\paragraph{About 1 epoch to sync, even for high fill rates.}
The results in Figure~\ref{fig:latency} are the average of three runs in each
drip rate.
Again, there are two general regimes to the graph, a linear one and a
super-linear one, and the transitions between them are, again, better than our
theoretic 25\% bound. First, for lower fill rates, the time to complete a single
file synchronization is {\em roughly one epoch}. 

At higher fill rates, it starts to take more epochs, on average, to sync a
single file; however, even for the most conservative $k=3$, it only takes at
most 5 epochs even for very high fill rates. For more aggressive drip rates,
$k=9,12$ the impact of higher filler rates is diminished, still only requiring
about 2 epochs to synchronize a single file. This is to be expected as selecting
more backend files for writing increased the likelihood of finding space to
clear the buffer.

\subsection{The Size of Pending Writes Buffer}

In this experiment, we investigate how much space the pending writes buffer
requires while the system is working. To do so, we consider more realistic file
sizes and file write patterns under high thrashing rates, contrary to most of
the previous experiments where each frontend file has two full-block fragments. 

We inserted frontend files of varied size based on known file size
distributions such that the backend was filled to 20\%, 50\%, or
75\%. The file sizes were based on prior published work in measuring
file size distributions. In particular, we followed a lognormal
distribution, which has been shown to closely match real file sizes
\cite{Dow01}, fit with data from a large-scale study of file sizes
used in a university setting \cite{THB06}. The same distribution was
used in the variable file size experiment previously.

We also generated a series of writes to these files such that, on average, 1MB
of data was updated on each write. This could occur within a single file or
across multiple files. We selected which file to thrash based on distributions
of actual write operations in the same study \cite{THB06}, used to
generate the original file sizes. Roughly, this distribution gives a stronger
preference to rewriting smaller files. We did {\em not write exactly} 1MB of
data in each batch, but rather kept the average of each batch size as close to
1MB as possible in accordance with the experimental write size distribution. In
particular, there were batches were a file larger than 1MB was written. 
As before, we used $N=1024$ and $B=1\mbox{MB}$. We used the drip rate $k=3$ to
show the most conservative setting of \sys. We averaged the results of three
independent runs.

\begin{figure}[t]
  \centering
  \includegraphics[width=\minof{10 cm}{0.95\linewidth}]{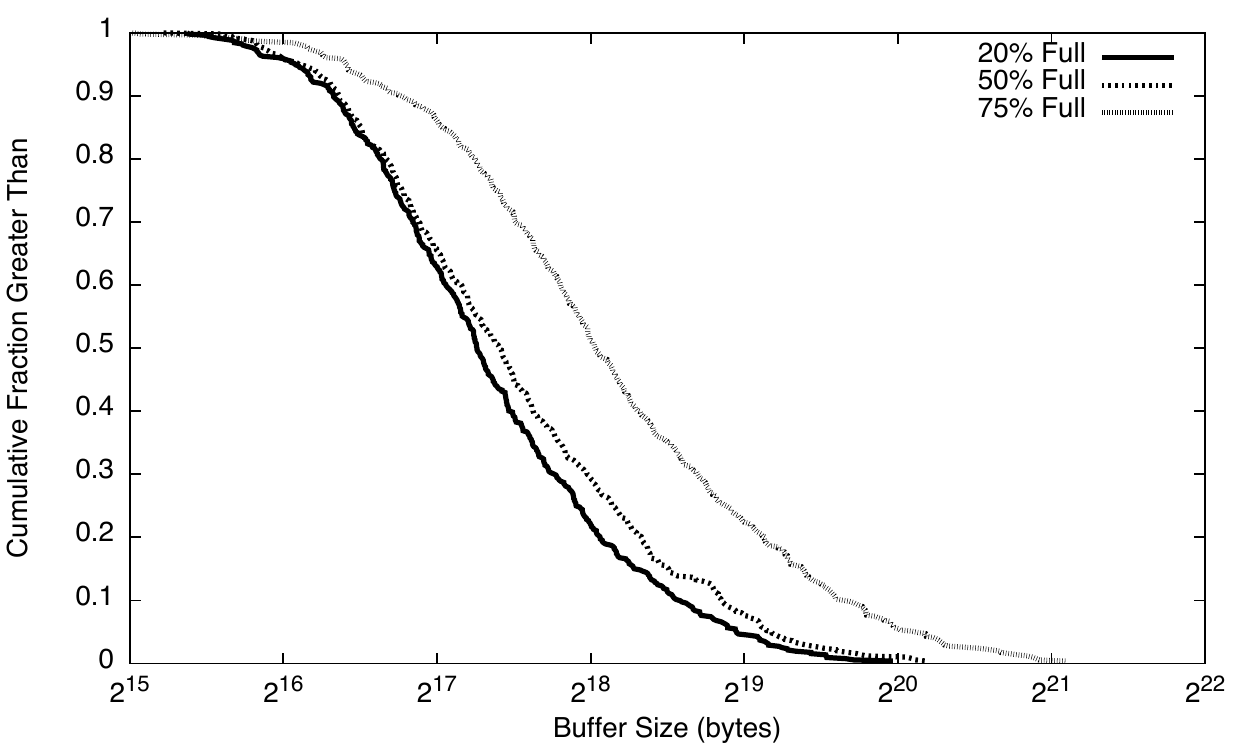}
  \caption{Buffer size under realistic file distributions. The experiment was
      performed with parameters $N=1024, B=1\mbox{MB}, k=3$. 
  The point $(x, y)$ means for $y$ fraction of observed execution time, the size
  of the buffer was greater than $x$. For example, when the backend is filled
  with 20\% (the solid line with 20\% full), only for 0.2 fraction of the entire
  execution time, the buffer size is roughly greater than $2^{18}$ bytes. In
  addition, the buffer size is always larger than $2^{15}$. 
  \label{fig:buf}
   }
\end{figure}

\paragraph{Reasonable buffer size: at most 2 MB.} 
The primary result is presented in Figure~\ref{fig:buf} where we measure the
number of bytes in the buffer on each synchronization. In the graph, the point
$(x, y)$ means for $y$ fraction of observed execution time, the size of the
buffer was greater than $x$. For example, when the backend is filled with 20\%
(the solid line), only for 0.2 fraction of the observed
execution time, the buffer size is roughly greater than $2^{18}$ bytes.  In
addition, the buffer size is always larger than $2^{15}$, and {\em the buffer
never grows larger than about 2 MB}, which corresponds to only 4 blocks.

Clearly, as the fill rate increases, the amount of uncommitted data in the
buffer increases; however, the relationship is not strictly linear. 
For example, with 20\% full and 50\% full, we see only a small difference in the
buffer size for this extreme thrashing rate. The synchronization is able to keep
up with the high thrashing rate for two main reasons: first, on each
synchronization, it is generally able to clear something out of the buffer; and
second, some writes occur on the same files and on small files (as would be the
case in a real file system), which allows these writes to occur on cached copies
in the buffer and the smaller files to packed together efficiently into blocks,
even partially full ones.

At a fill rate of 75\%, however, there is a noticeable performance degradation.
Because most of the blocks selected at each epoch are either full or do not have
enough space, due to fragmentation, the buffer cannot always be cleared at a
rate sufficient to keep up with incoming writes. Thus, the size of the buffer
doubles in comparison with the other workloads.

\subsection{Measurements on Dropbox}
Here, we measure the performance of \sys on a real cloud synchronization
service, namely Dropbox. 

We performed both a latency and throughput measurement,
just as before, using 1MB backend files, but this time the backend directory was
stored within a Dropbox synchronized folder with measurements taken across two
computers on our institution's network that had access to the synchronized
folder.

In these experiments, we used a drip rate of $k=3$ and a drip time
$t=10$ seconds. We experimented with lower drip times, but found that due to
rate limiting of the Dropbox daemon on the synchronized folders, a longer drip
time is required in order to to ensure that we do not blow through the epoch
boundary.

In both the latency and throughout experiments described below, we
established a dropbox connection on two computers on our institution's
network. We designated one computer as the writer and one the
reader. On the writer, it ran \sysrw and \sysro, on a shared backend
folder stored within the Dropbox synchronization folder. The writer
computer measured the amount of time it took for the \sysrw to
synchronize to the local \sysro mount. Meanwhile, the reader computer
ran a \sysro mount only and monitored the file systems for the
appearance of synchronized files. The difference between the \sysro
mount on the write computer and the \sysro mount on the read computer is
the propagation delay imposed by Dropbox. Additionally, as we do not
have insight into how the Dropbox daemon chooses to synchronize, it is
also possible that other factors are coming into play, such as taking
incremental snapshots of deleted files.

\paragraph{Baseline: EncFS.}
Additionally, we wish to provide a baseline comparison of the overhead
of \sys, and so we performed similar experiments using
EncFS~\cite{encfs} as the data protection mechanism. Much like \sys,
EncFS using a FUSE mount to display an encrypted file system that is
stored in a backend folder, but EncFS provides no oblivious
protection. Instead files are simply stored individually encrypted, so
the total number of files is revealed as well as their 
sizes and full access patterns.

\begin{figure}[t]
  \centering
  \includegraphics[width=\minof{10 cm}{0.95\linewidth}]{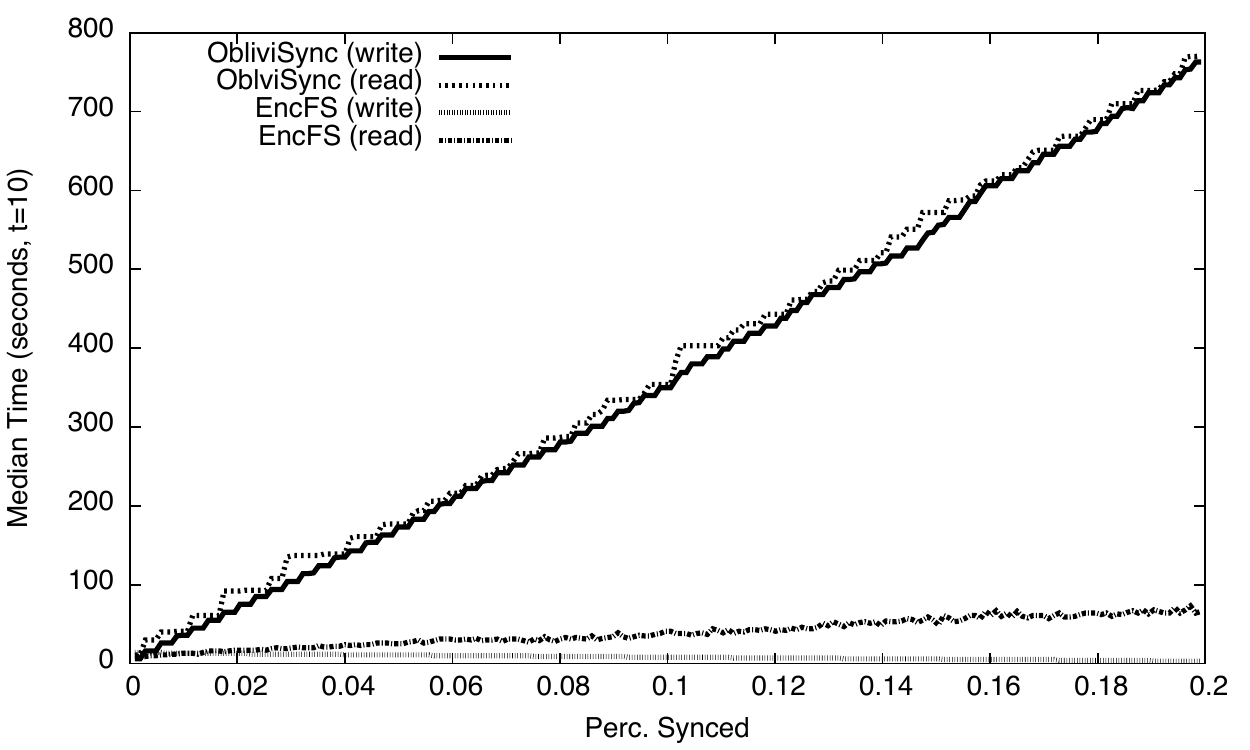}
  \caption{Throughput of writing 0.25GB of 1MB files all at once for
    \sys and EncFS on Dropbox synchronized backend between two
    machines running on the same network.  We used a drip time of 10
    seconds for \sys ($t=10$) and a drip rate of 3 ($k=3$) for a
    conservative estimate.}
  \label{fig:dbox:throughput}
\end{figure}

\paragraph{Throughput over Dropbox.}
The throughput measurement occurred much like as described earlier in
the section. For both EncFS and \sys, we inserted a large number of
files, namely 20\% full or $\sim$200MB, and then we measured how long
it takes for the buffer to clear and all files to become
available. Like before, we used a read and write computer, and we
measured the difference in the local and remote propagation delays of
file synchronization. The primary result is presented in
Figure~\ref{fig:dbox:throughput}.

For EncFS on the write computer, the propogation delay for all
the files is nominal with files appearing nearly immediately. On the
read computer, there is a propagation delay associated with Dropbox
remote synchronization, and all files are accessible within 100
seconds.
For \sys on the write computer, we see a very similar throughput
trend-line as in the prior experiments. In total, it takes just under
800 seconds (or 80 epochs) for all the files to
synchronize. Interestingly, on the read computer, the propagation
delay is relatively small, with respect to the overall delay, and
files are accessible within an additional epoch or two.
{\em In total, these results clearly demonstrate that \sys is functional and
efficient to use over cloud synchronization services like Dropbox.}

\begin{figure}[t]
  \centering
  \includegraphics[width=\minof{10 cm}{0.95\linewidth}]{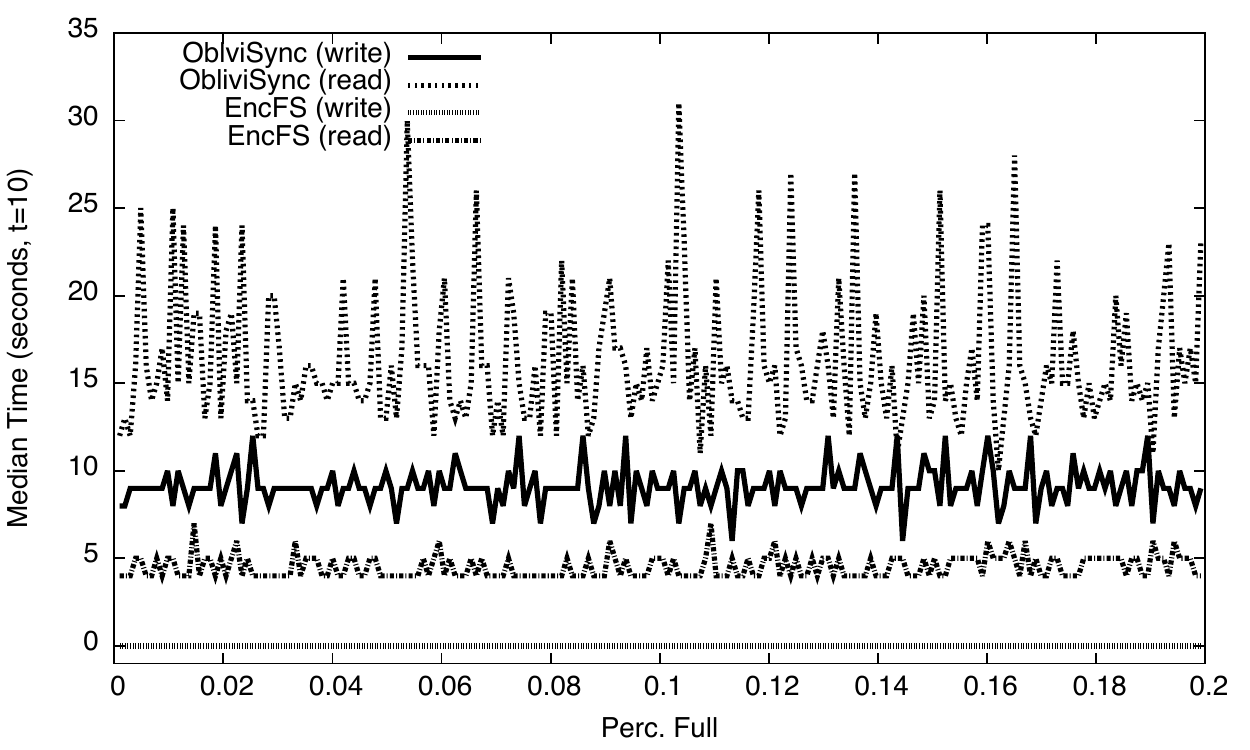}
  \caption{Latency of writing 1MB files one at a time for \sys and
    EncFS on Dropbox synchronized backend between two machines running
    on the same network.  We used a drip time of 10 seconds for \sys
    ($t=10$) and a drip rate of 3 ($k=3$) for a conservative
    estimate. The variations in the trend lines are likely due to
    jigger in the network.}
  \label{fig:dbox:latency}
\end{figure}

\paragraph{Latency over Dropbox.}
Figure~\ref{fig:dbox:latency} shows the primary result of running \sys
and EncFS using Dropbox as the cloud synchronization service. The
EncFS write line is nearly 0 (s) as immediately upon writing the file
it becomes available to write computer. However on the read computer,
it takes a little under 5 seconds for the synchronization with Dropbox
to complete for the same file to be accessible. This measurement forms
a baseline of performance for the rate of DropBox
synchronization without \sys.

For \sys, on the write computer, we see an expected performance metric
of just under 10 seconds for each file to be visible to the read mount. The
reason it is under 10 seconds and not exactly 10 seconds, as the
setting of the drip time, is that a write occurring between epoch
timers will take less than an epoch to sync. The propagation rate to
the read computer takes a similar time as that of EncFS ($\sim$ 5
seconds); however, there is higher variance as more files need to be
transferred by the Dropbox service per epoch (namely $4=k+1$ with the
superblock).  Still, this added variance is within 3x in terms of
epochs: it takes at most 30 seconds for a file to sync (or 3 epochs of
waiting), which is very reasonable considering the built-in overhead of
the system.

\section{Related Work}

\paragraph{ORAM.} ORAM protects the access pattern from an observer such that it
is impossible to determine which operation is occurring, and on which
item. The seminal work on the topic is by Goldreich and
Ostrovsky~\cite{GO96}, and since then, many works have focused on
improving efficiency of ORAM in both the space, time, and communication
cost complexities (for example \cite{AC:SCSL11,kushilevitz2012security,CCS:SDSFRY13,usenix:DSS14,CCS:MoaMayBla15}
just to name a few; see the references therein).
Blass~et~al.~introduced write-only ORAMs~\cite{CCS:BMNO14}. In a
write-only ORAM, any two write accesses are indistinguishable, and they
achieved a write-only ORAM with optimal $O(1)$ communication complexity
and only poly-logarithmic user memory. 
Based on their work, we construct a write-only ORAM that additionally
supports variable-size data and hides the when the data items are
modified. We point out also that variable-sized blocks in traditional
read/write ORAMs were also considered recently by \cite{RAC16}, but with
higher overhead than what can be achieved in the write-only setting.

\paragraph{Protecting against timing side-channels.}
Side-channel attacks that use network traffic analysis in order to learn
private information have been considered in contexts other than secure
cloud storage. Proposed systems for location tracking \cite{RMKK08} and
system logging \cite{BHJT14} use buffering and random or structured
delays to protect against such attacks in a similar way to our work.

\paragraph{Personal cloud storage.} 
A personal cloud storage offers automatic backup, file synchronization,
sharing and remote accessibility across a multitude of devices and
operating systems.  
Among the popular personal cloud storages are Dropbox, Google Drive, Box,
and One Drive. 
However, privacy of cloud data is a growing concern, and to address this
issue, many personal cloud services with better privacy appeared. Among
the notable services are SpiderOak~\cite{spideroak},
Tresorit~\cite{tresorit}, Viivo~\cite{viivo},
BoxCryptor~\cite{boxcryptor}, Sookas~\cite{sookasa},
PanBox~\cite{panbox}, and OmniShare~\cite{omnishare}.  
All the solutions achieve better privacy by encrypting the file data
using encryption keys created by the client. 
We stress that however there has been no attempt to achieve the stronger
privacy guarantee of obliviousness.

\section{Conclusion}

In this paper, we report our design, implementation, and evaluation of
\sys, which provides oblivious synchronization and backup for the
cloud environment. Based on the key observation that for many cloud
backup systems, such as Dropbox, only the writes to files are revealed
to cloud provider while reads occur locally, we built upon write-only
ORAM principles such that we can perform oblivious synchronization and
backup while also incorporating
protection against timing channel attacks. When the drip-rate and drip
time parameters are set properly according to the usage pattern, this
overhead is just 4x both in theory and in practice.

We also consider practicality and usability. \sys is designed to
seamlessly integrate with existing cloud services, by storing
encrypted blocks in a normal directory as its backend.  The backend
can then be stored within any cloud based synchronization folder, such
as a user's Dropbox folder. To be stored within the backend encrypted
blocks, we designed a specialized block-based file system that can
handle variable size files. The file system is presented to the user
in a natural way via a frontend FUSE mount such that the user-facing
interface is simply a folder, similar to other cloud synchronization
services. Any modifications in the frontend FUSE mount are
transparently and automatically synchronized to the backend without
leaking anything about the actual writes that have occurred.

In evaluating our system, we can prove that the performance guarantees
hold when 25\% of the capacity of the backend is used, and our
experimental results find that, with realistic workloads, much higher
capacities can in fact be tolerated while maintaining very reasonable
efficiency. Importantly, \sys can be tuned to the desired application
based on modifying the drip rate and drip time to meet the
application's latency and throughput needs.

Although \sys works well in practice already, there are still
interesting and difficult open problems in this domain. While we have
optimized the efficiency for the \emph{client}, cloud service providers
may be hesitant to encourage systems such as \sys because they will
eliminate the possibility of \emph{deduplication} between users, where
common files are stored only once by the service provider. Furthermore,
as our system only allows one \sysrw client at any given time, an
important use-case of \emph{collaborative editing} is not permitted
here. It may be necessary to overcome challenges such as these 
in order to bring oblivious cloud storage into mainstream use.

\section*{Acknowledgements}
This work was supported in part in part by ONR awards N0001416WX01489 and
N0001416WX01645, and NSF award \#1618269, \#1406177, and \#1319994.


\end{document}